\documentclass[iop]{emulateapj}
\usepackage{graphics,enumerate,amsmath, morefloats}
\usepackage[utf8]{inputenc}
\usepackage[T1]{fontenc}
\usepackage{lmodern}

\newcommand{\heiilw}{\mbox{\ion{He}{2} $\lambda$4686}}

\newcommand{\oiiw}{\mbox{[\ion{O}{2}] $\lambda$3727}}

\newcommand{\niibw}{\mbox{[\ion{N}{2}] $\lambda$6583}}
\newcommand{\niiw}{\mbox{[\ion{N}{2}] $\lambda \lambda$6548,6583}}

\newcommand{\oiiiw}{\mbox{[\ion{O}{3}] $\lambda \lambda$4959,5007}}

\newcommand{\oii}{\mbox{[\ion{O}{2}]}}
\newcommand{\nii}{\mbox{[\ion{N}{2}]}}
\newcommand{\hal}{\mbox{H$\alpha$}}

\newcommand{\hb}{\text{H$\beta$}}

\shorttitle{Flux Calibration for MaNGA}
\shortauthors{Yan et al.}

\begin{document}
\title{SDSS-IV/MaNGA: Spectrophotometric Calibration Technique}
\author{Renbin Yan$^{1}$, Christy Tremonti$^{2}$, Matthew A. Bershady$^{2}$, David R. Law$^{3}$, David J. Schlegel$^{4}$, Kevin Bundy$^{5}$, Niv Drory$^{6}$, Nicholas MacDonald$^{7}$, Dmitry Bizyaev$^{8,9}$, Guillermo A. Blanc$^{10,11,12}$, Michael R. Blanton$^{13}$, Brian Cherinka$^{14}$, Arthur Eigenbrot$^{2}$, James E. Gunn$^{15}$, Paul Harding$^{16}$, David W. Hogg$^{13}$, Jos\'e R.~S\'anchez-Gallego$^{1}$,  Sebastian F. S\'anchez$^{17}$, David A. Wake$^{2,18}$, Anne-Marie Weijmans$^{19}$, Ting Xiao$^{20}$, Kai Zhang$^{1}$}

\affil{$^1$ Department of Physics and Astronomy, University of Kentucky, 505 Rose St., Lexington, KY 40506-0057, USA; yanrenbin@uky.edu}
\affil{$^2$ Department of Astronomy, University of Winsconsin-Madison, 475 N. Charter Street, Madison, WI 53706-1582, USA}
\affil{$^3$ Space Telescope Science Institute, 3700 San Martin Drive, Baltimore, MD 21218, USA}
\affil{$^4$ Physics Division, Lawrence Berkeley National Laboratory, Berkeley, CA 94720-8160, USA}
\affil{$^5$ Kavli IPMU (WPI), UTIAS, The University of Tokyo, Kashiwa, Chiba 277-8583, Japan}
\affil{$^6$ McDonald Observatory, Department of Astronomy, University of Texas at Austin, 1 University Station, Austin, TX 78712-0259, USA} 
\affil{$^7$ Department of Astronomy, Box 351580, University of Washington, Seattle, WA 98195, USA}
\affil{$^8$ Apache Point Observatory, P.O. Box 59, Sunspot, NM 88349, USA} 
\affil{$^9$ Sternberg Astronomical Institute, Moscow State University, Universitetskij pr. 13, Moscow, Russia}

\affil{$^{10}$ Departamento de Astronomía, Universidad de Chile, Camino el Observatorio 1515, Las Condes, Santiago, Chile}
\affil{$^{11}$ Centro de Astrofísica y Tecnologías Afines (CATA), Camino del Observatorio 1515, Las Condes, Santiago, Chile}
\affil{$^{12}$ Visiting Astronomer, Observatories of the Carnegie Institution for Science, 813 Santa Barbara St, Pasadena, CA, 91101, USA}

\affil{$^{13}$ Center for Cosmology and Particle Physics, Department of Physics, New York University, 4 Washington Place, NY 10003, New York}
\affil{$^{14}$ Dunlap Institute for Astronomy and Astrophysics, University of Toronto, 50 St. George Street, Toronto, Ontario M5S 3H4, Canada}
\affil{$^{15}$ Department of Astrophysical Sciences, Princeton University, Princeton, NJ 08544, USA} 
\affil{$^{16}$ Department of Astronomy, Case Western Reserve University, Cleveland, OH 44106, USA}
\affil{$^{17}$ Instituto de Astronomia, Universidad Nacional Autonoma de Mexico, A.P. 70-264, 04510 Mexico D.F., Mexico}
\affil{$^{18}$ Department of Physical Sciences, The Open University, Milton Keynes, MK7 6AA, UK}
\affil{$^{19}$ School of Physics and Astronomy, University of St Andrews, North Haugh, St Andrews KY16 9SS, UK}
\affil{$^{20}$ Shanghai Astronomical Observatory, Nandan Road 80, Shanghai 200030, China}

\begin{abstract}
Mapping Nearby Galaxies at Apache Point Observatory (MaNGA), one of three core programs in the Sloan Digital Sky Survey-IV (SDSS-IV), is an integral-field spectroscopic (IFS) survey of roughly 10,000 nearby galaxies. It employs dithered observations using 17 hexagonal bundles of 2\arcsec\ fibers to obtain resolved spectroscopy over a wide wavelength range of 3,600-10,300\AA. To map the internal variations within each galaxy, we need to perform accurate {\rm spectral surface photometry}, which is to calibrate the specific intensity at every spatial location sampled by each individual aperture element of the integral field unit. The calibration must correct only for the flux loss due to atmospheric throughput and the instrument response, but not for losses due to the finite geometry of the fiber aperture. This requires the use of standard star measurements to strictly separate these two flux loss factors (throughput versus geometry), a difficult challenge with standard single-fiber spectroscopy techniques due to various practical limitations. Therefore, we developed a technique for spectral surface photometry using multiple small fiber-bundles targeting standard stars simultaneously with galaxy observations. We discuss the principles of our approach and how they compare to previous efforts, and we demonstrate the precision and accuracy achieved. MaNGA's relative calibration between the wavelengths of \hal\ and \hb\ has a root-mean-square (RMS) of 1.7\%, while that between \niibw\ and \oiiw\ has an RMS of 4.7\%. Using extinction-corrected star formation rates and gas-phase metallicities as an illustration, this level of precision guarantees that flux calibration errors will be sub-dominant when estimating these quantities. The absolute calibration is better than 5\% for more than 89\% of MaNGA's wavelength range. 
\end{abstract}

%This paper just talks about the fluxcal technique. 
%We do not discuss the selection of standard stars, or we assign bundles to them. These will be included in the survey paper. 

\section{Introduction}\label{sec:intro}%review of spectrophotometry calibration methods used by other surveys.

Spectrophotometry refers to the calibration of the observed flux density as a function of wavelength to the intrinsic
flux density of the target. This calibration is critically important for deriving accurate quantities for many physical
properties from spectroscopic measurements of galaxies, including emission line measures of star formation rates and
gas-phase metallicities and stellar population parameters from spectral fitting. The success of Sloan Digital Sky
Survey \citep{York00} would not be possible without its accurate spectrophotometric calibration. In SDSS-I, -II and
-III, multiple standard stars were observed simultaneously with the science targets, and the achieved calibration
accuracy is on the order of 5\% \citep{SDSSDR6, Dawson13}.

%standard stars are observed simultaneously with galaxy targets. SDSS surveys have used fibers inserted in plug plates to target galaxies and standard stars simultaneously. Each plate includes one or two dozen standard stars spread all over the field. This allowed the correction for the imperfect throughput of the atmosphere, telescope, and instrument, as well as the correction for the different aperture loss due to drilling error and differential atmosphere refraction. The resuling absolute calibration for SDSS-I and SDSS-II is accurate to better than 10 percent and the relative calibration between different wavelength is accurate to a few percent. For SDSS-III/BOSS, the calibration accuracy is slightly worse due to the use of fainter standard stars. 

%Introduction to MaNGA. (one paragraph).
The MaNGA (Mapping Nearby Galaxies at Apache Point Observatory) project \citep{Bundy15} is an integral field spectroscopic (IFS) survey of nearby galaxies using the 2.5-m Sloan Foundation Telescope \citep{Gunn06} and the BOSS spectrographs \citep{Smee13}. It is one of three surveys that comprise SDSS-IV, which started in July 2014. With 17 hexagonal fiber bundles \citep{Drory15}, deployed across each 3 degree diameter pointing, MaNGA will obtain spatially-resolved spectroscopy for roughly 10,000 nearby galaxies by 2020. The fiber bundles are made with 2\arcsec\ fibers and have sizes ranging from 12\arcsec\ to 32\arcsec\ diameter in the long axis. The spatial fill factor is 56\%.  The two BOSS spectrographs, each with a blue and a red camera, provide a wavelength coverage from 3,600\AA\ to 10,300\AA\ at a resolution of $R\sim2000$.

Different from other previous and current SDSS surveys that target each source with only one fiber, MaNGA will cover and map individual galaxies. This important difference reshapes the
goal of spectrophotometry in the IFS context.  For MaNGA, we wish to calibrate spectral \emph{surface} photometry as
we explain below.
%desire a different approach to calibration based on specific intensity rather than flux, i.e., our requirement is to calibrate spectral surface photometry. 

In spectroscopic studies of external galaxies, stars have always been used as calibrators for spectrophotometry. However, stars are effectively point sources, while external galaxies often appear as extended sources and in the MaNGA sample cannot be approximated as point sources. Because of this difference between the calibrator and the object of study, the detailed approach of spectrophotometry varies depending on the particulars of the instrument and observation setup, and the desired goal of the calibration.

When the spectroscopic aperture is much larger than the size of the point spread function (PSF) at all relevant wavelengths, flux calibration
using a star can be a trivial exercise. When the aperture is smaller or comparable to the size of the PSF, some
fraction of the light from a point source will fall outside the aperture and be lost, with the amount of loss depending on the location of the source within the aperture.  Usually, instrument apertures are
more closely matched to the PSF for the sake of maximizing the obtained signal-to-noise ratio and optimizing spectral
resolution. However, apertures placed on an extended source will not see the same amount of flux loss as for point
sources for the simple reason that as some light is shifted out of the aperture other light may be shifted in. The
exact amount of light either lost or gained in this manner as the effective location of the aperture changes will be a complicated function of the 2D surface brightness profile of the target.  In such cases, there are at least three different spectrophotometry goals as applied to galaxy targets.
%When the aperture is smaller or comparable to the size of the PSF, the flux loss (or gain) factor can be different between a point source and an extended source. The latter case is more commonly adopted for the sake of maximizing signal-to-noise ratio and optimize spectral resolution. 
%In these cases, there are at least three different desired spectrophotometry goals. 

%In the latter case, a reasonable goal of spectrophotometry is to calibrate the slit- or fiber-aperture flux of a PSF-convolved spatial profile as observed above the atmosphere. 

\begin{enumerate}[A.]
\item Calibrate to the slit- or fiber-aperture flux density ($f_\lambda$) of a PSF-convolved spatial profile, or in
other words, the specific intensity (a.k.a.~surface brightness) integrated within the measurement aperture of a
PSF-convolved spatial profile. Here the PSF includes the combined effects of atmospheric seeing, the point spread
function of the telescope and instrument, and chromatic aberration in the whole system.  The goal is to correct for the
atmospheric attenuation of the flux density and the instrument response, but not to deconvolve the PSF or correct for
geometric shifts due to differential atmospheric refraction (DAR).

\item Calibrate to the total flux density incident on the atmosphere if the galaxy were a point source. This is in
practice straightforward because the same flux correction vector is applied to both stars and galaxies. But it assumes the target galaxies experience the same DAR and aperture flux losses as the stars do, which is usually not true.

\item Calibrate to the total flux density derived from imaging photometry assuming that the relative shape of the spectral energy distribution is uniform within the galaxy. The uniformity assumption is appropriate only for certain science cases.

\end{enumerate}

We consider the first of the above options the most fundamental goal for spectrophotometry. It truly reflects what is
being measured. It makes no assumption about the property of the extended source to be observed. The only correction
required is the system throughput, without any flux correction due to geometric factors. However, this goal is
difficult to achieve given practical limitations, especially for single-fiber spectroscopy, as we will detail below.
For slit spectroscopy, one approach is to place a slit much wider than the PSF on standard stars to obtain the needed
correction, with the caveat that the resulting spectral resolution will be different.

Given the difficulty of actually achieving Goal A, many observational projects have chosen to fall back to Goal B or C. For single-fiber spectroscopy of galaxies, especially distant ones where galaxies are marginally resolved, these can be sufficient for the purpose of deriving redshifts and measuring approximately-global spectral properties. 

However, in the IFS context, the ultimate goal is to study the internal variations within a galaxy. Therefore, Goal A
is the only sensible choice for spatially mapping the specific intensity as a function of
wavelength. There are a number of practical difficulties, however, which we discuss in detail in this paper. For MaNGA, we have developed and tested a method to achieve this goal. The approach we present here is broadly applicable to other IFS studies of extended sources.

This paper is organized in the following way. In Section 2, we first discuss the causes of flux loss and error, how spectrophotometry was done in previous generations of SDSS, and the different spectrophotometry needs for integral field spectroscopy. 
%describe how spectrophotometry is done for single fiber spectroscopy, specifically in previous generation of SDSS, then we describe how IFU spectrophotometry is different and why it requires a different technique. 
In Section 3, we discuss how we set the requirements for spectrophotometry given the MaNGA science requirements. 
%In Section 4, we discuss the various calibration methodologies we have considered. 
We then describe our calibration method and the implementation in Section 4, present the resulting spectrophotometry accuracy achieved in Section 5 and summarize in Section 6.

\section{What to correct: sources of flux errors}\label{sec:principle}
\subsection{Sources of flux loss and flux error}

%In this section, we discuss how spectrophotometry calibration is done for single-fiber spectroscopy from SDSS-I to SDSS-III. There are many issues that are similar as in SDSS-IV/MaNGA. To understand how to do the calibration, we have to first understand what are the various causes for flux error. There are mainly two categories of causes. 
To evaluate whether a spectrophotometric calibration method will achieve the above Goal A, we first have to understand the various reasons why observed spectra differ from the intrinsic spectra of the targets.
We put these flux losses and erorrs into two categories. 

\subsubsection{Throughput loss}
%\item{\bf Throughput loss}

The first is flux loss due to imperfect throughput of the system, including atmospheric transparency, reflectance and transmission of all optical elements in the telescope and instrument (including fibers), and CCD quantum efficiency. All these throughput losses are a function of wavelength. %The difference in throughput among different fibers are corrected using spectroscopy flat-field frames. The other throughput factors are basically the same for alll fibers in the system. 

\subsubsection{Aperture-induced flux error}

The second kind of flux error is due to aperture mis-centering which can also lead to wavelength-dependent flux errors.
 We refer to this as flux error rather than flux loss because for extended sources unaccounted flux can be both added
or lost. The list of causes of this kind of flux error differs for point sources and extended sources.
Common to both are mechanical alignment errors from manufacturing, guiding errors at the guiding wavelength, and DAR. In detail, the exact source of these errors and their significance depend on the performance of the observing system hardware and the observing strategy. Below, for the specific case of SDSS, we go through each source in detail. 
% given the difference in their spectrophotometry goals.

%depend on the goals of the spectrophotometry, thus it is different for point sources and extended sources. %We discuss this separately.
%This aperture-related flux error is different for a point source and for an extended source because the goals of the spectrophotometry is different. 

%For point sources, we can define the flux error as the difference between the observed flux density ($f_\lambda$ or $f_\nu$) and the total flux density enclosed in the PSF. For extended source much larger than the aperture which could potentially have internal variation, one reasonble definition for the flux error is the difference between the observed flux density and the aperture flux density of a PSF-convolved surface brightness profile. Below, we discuss these two cases separately.

%For extended sources, 

\begin{enumerate}
\item \textsf{Fiber Positioning:} In SDSS, fibers are positioned on science targets by being plugged into custom-drilled aluminum plates that are
mounted at the telescope's focal plane. The holes on the plug plates have positional errors from drilling. The fibers
are held within their indiviual metal housings (so called ferrules), which are plugged into the holes. The fiber is not
always perfectly centered within the ferrule due to limited precision in manufacturing. The plate hole needs to be
slightly larger than the ferrule in order for it to be pluggable, and as a result the ferrule will not be perfectly
centered within the hole either. The fiber centration error within the ferrule, the hole-ferrule clearance, and the
positional error from drilling can stack up to 0\farcs36 root-mean-square positional error on the target (see
\citealt{Drory15} for the detailed error stack up), as compared to the 2\arcsec\ diameter fibers used in SDSS-III and
IV. The dominant component is the drilling error. A large part of the drilling offset can be measured post-drilling,
and in principle could be taken into account in the spectrophotometric calibration. In practice, this was not done in
previous generations of SDSS as it was not deemed scientifically essential.

\item \textsf{Monochromatic Atmospheric Field Distortions:} The monochromatic component of the atmospheric refraction (AR) distorts the field in a non-circularly-symmetric
way when the telescope is not pointed at zenith. When a plate is drilled, the offsets due to AR at the guide wavelength
are taken into account according to the hour angle and altitude at which the plate is planned to be observed. However,
observations can last several hours during which the magnitude and direction of the AR will change causing a misalignment between the fiber and the target. Given the Sloan Telescope's wide 3\arcdeg{} diameter field-of-view, the misalignment can be signficant.
By tuning the distance between the primary and the secondary mirror, the scale of the field can be adjusted to partially compensate. However, the quadrupole distortion cannot be corrected (for more details, see Sec 4.2 of \citealt{Law15}). This means some fibers, depending on their positions on the plate,  will be offset from the target even if guiding is perfect. The global guiding error for SDSS is expected to be much smaller than all these effects. 

For example, at a zenith distance of 18\arcdeg{} (airmass of 1.05), the compression of the 3\arcdeg{} field in the
altitude direction is 2\farcs4. Compensating with the scale change, the residual offset due to the AR for a target on
the plate could be somewhere between 0-0\farcs6 at the guiding wavelength. The global guiding error is on the order
of 0\farcs05.

\item \textsf{Differential Atmospheric Refraction:} The third contributor to the aperture centering error is the {\it
differential} atmosphere refraction. This means the images of the targets at blue wavelengths are offset from those at
red wavelengths. At an airmass of 1.05, the separation between the monochromatic images at 3600\AA\ and 10,300\AA\ is
0\farcs54. At airmass 1.25, it is 1\farcs27. For a point source, this means the flux loss due to a finite fixed aperture is different for different wavelengths (e.g., a point source centered in a fiber at one wavelength may fall near the edge of that fiber at another wavelength). For an extended source, this means the fiber is seeing different parts of the source at different wavelengths. The spectrum one eventually extracts from an individual fiber contains mixed information from different parts of the galaxy. In slit spectroscopy, one could align the slit with the parallactic angle to capture all the flux. For single-fiber spectroscopy on extended sources with internal variations, we will not be able to correct for DAR to get a spectrum for the same physical aperture at all wavelengths, because we cannot correct for flux that we do not observe and is a priori unknown. This is why we excluded DAR corrections in Goal A above, and why Goal A is the most sensible spectrophotometry goal for extended sources. 

In SDSS-I to SDSS-III, the approach of Goal B was adopted for spectrophotometry. Due to the different flux loss
experienced by point sources and extended sources, there could be significant wavelength-dependent systematics in the
flux calibration for each galaxy, especially when DAR is large. For many science topics this may not matter, but
avoiding such systematics becomes critical in the context of IFS.
%When Goal B is adopted, the implicit assumption is that the light is dominated by the centra
%to interpret the galaxy spectra, an implicit assumption has to be made, which is that the spectral shape is uniform in each target galaxy. With this assumption, we could achieve Goal B or C, or something similar. This is sufficient for many science topics, but not for science targeted by IFS.
%to interpret the spectra one has to assume that the property of the target galaxy is uniform on this scale, except for variation in overall normalization. Here, the corrections needed for stars and for galaxies are different.

\item \textsf{Seeing and Chromatic Aberrations:} For point sources, aperture losses arise from  two additional factors,
both of which lead to wavelength-dependent PSF variation. The first is the wavelength-dependent seeing profile. The
second is the chromatic aberration of the system. For example, for the Sloan Telescope, the plate is designed to follow
the focal plane shape at 5300\AA. The focal planes for other wavelengths are different.  The resulting PSF shape as a
function of wavelength as seen by fibers at different plate locations can be distorted. 

The treatment of these effects for extended sources depends on the spectrophotometry goals. For example, for Goal A,
these two factors should be included in the intrinsic source properties for which there should be no corrections. What one observes with fiber spectroscopy is the aperture flux of the surface brightness distribution convolved with the wavelength-dependent PSF. One cannot reliably deconvolve the PSF without knowing the intrinsic intensity distribution within each galaxy. If, on the other hand, one adopts Goal B for practical reasons, then galaxies are assumed to experience the same flux loss due to these two factors as stars do, even though this assumption is in general incorrect.
%One may argue these two latter factors also matter for extended sources. 
%For extended sources with intrinsic flux variations on spatial scales smaller than the PSF, such as nearby galaxies targeted by SDSS-I, these two factors shall be treated as part of the ``truth''. 

\end{enumerate}

%The error in fiber centering is largely unknown. But the majority of them  
The first three factors above are all related to alignment.  Their combined effect are different for stars and
galaxies. For stars, a certain fraction of flux is lost as a function of wavelength and the needed correction factor is
usually a slow function of wavelength.  There are no high-frequency changes to the spectral shape. For galaxies, the
impact is more complicated because alignment errors combined with DAR mean that different parts of the galaxy are
sampled at different wavelengths.

%The combined effect of these fiber alignment errors causes a flux scaling error as a function of wavelength and position on the plate. The exact magnitude is unknown due to the largely unknown fiber-centering error. 

%\bigskip

%In addition to the above sources that are common to both point sources and extended sources, there are two other sources that shall be treated differently for point and extended sources. 

%\bigskip

Given the sources of flux errors above, it is clear that IFS requires calibration of spectral surface photometry (i.e.,
Goal A), which necessitates corrections only for the throughput loss of the system but not any aperture-induced flux
error. However, because we use stars as calibrators, they do experience aperture-induced flux error as well. 
Thus, to separate these two sources of flux errors for calibration stars, we have to know exactly how the stars are positioned relative to the spectroscopic aperture and the shape of the PSF. 

\subsection{Calibration for Single-fiber Spectrsocopy in SDSS-III/BOSS}\label{sec:oldmethod}

Below we describe the flux calibration method used in SDSS-III/BOSS, since MaNGA is using the same spectrographs and the same fiber size as BOSS. In SDSS-III/BOSS, 20 single fibers per plate were placed on standard stars. They were observed simultaneously with all the science targets. The light from the standard stars experienced the same throughput loss as the science fibers, with a small dependence on airmass. However, every fiber has a different aperture-induced flux error, due to their slightly different alignment error from manufacturing, drilling, and guiding, which are also compounded with DAR.

%The throughput loss are mostly the same for all fibers with a small dependence on airmass. %The PSF variation are also nearly the same with small dependence on airmass. 
%Their spectra also have flux error due to aperture-centering errors. However, as discussed above, the aperture-induced flux error are different for point sources and extended sources. 

The observed standard star spectra are first continuum-normalized using a running median filter with a width of 99
pixels ($\sim 110{\rm \AA}$ in the blue camera and $\sim140{\rm \AA}$ in the red camera) and then compared with a grid
of continuum-normalized Kurucz stellar models with different surface temperature, metallicity ([Fe/H]), and surface
gravity to find the best fitting models to all standards on the plate.  For each standard, a version of the chosen
model which has not been continuum-normalized is reddened using the extinction map of \cite{SchlegelFD98} and the
extinction law of \cite{O'Donnell94} and then scaled to match the $r$-band PSF magnitude of the star from its SDSS
imaging photometry. The calibration pipeline then compares the observed spectra of each plate's standard stars with the
reddened and normalized model spectra to determine a set of correction vectors. These corrections account for both the
throughput loss and the aperture-induced flux errors experienced by point sources.  Applying these corrections to a
galaxy is basically treating galaxies like point sources, what we refer to as Goal B spectrophotometry in Section
\ref{sec:intro}.

%Before SDSS Data Release 6 (DR6), the model spectra were scaled to match the 3\arcsec-fiber magnitudes of the stars \citep{SDSSDR6}. Here, the fiber magnitudes reflect only the flux contained within the aperture of a spectroscopic fiber. In DR6 and later releases including those for SDSS-III/BOSS (starting in DR9), the model spectra were instead scaled to match the PSF magnitudes of the stars from imaging. Conceptually, this is a switch from something close to Goal A to Goal B for spectrophotometry. The difference between PSF and fiber magnitudes for stars approximately reflects the aperture-induced flux error. If the model spectra are scaled to match fiber-magnitudes, the difference between the model and the data should just contain the throughput loss, in the ideal case without alignment error, guiding error, DAR, etc. This is essentially Goal A. However, additional steps in the pre-DR6 pipeline tries to correct for DAR, so it deviates the pre-DR6 results from Goal A. Post-DR6, the spectrophotometry goal switched to Goal B, which is to calibrate to the flux density as if galaxies were point sources. The correction vector derived in this case includes both the throughput loss and the zeroth-order term of the aperture-induced flux error.

Different alignment errors yield different aperture-induced flux errors among a plate's standard stars, leading to
correction vectors with significant differences in their overall shapes. First, the low-order shape difference is taken
out by dividing each correction vector by a cubic polynomial fit to their ratio to the mean correction vector. Then all
these low-order flattened correction vectors for all stars are combined together to derive an `average'
wavelength-dependent and airmass-dependent correction vector. Then the pipeline chooses a `best' exposure and corrects
the spectra from all the other exposures to match those in the best exposure on an object-by-object basis. This step is
required before all exposures can be coadded and involves only low-order polynomial scaling as a function of
wavelength.  Hence it can remove low-order flux differences caused by different DAR and guiding effects between multiple exposures.

Finally, after all exposures are combined, the pipeline solves for a flux distortion factor to correct for any
remaining flux error by comparing synthesized magnitudes from spectra with PSF magnitudes for stars and PSF-equivalent
magnitudes for galaxies. Using all galaxy and star targets, the code solves for a low order function that depends on
wavelength and plate position for each spectrograph. If the drilling error, guiding error, and DAR can all be
approximated by low order functions of wavelength and/or plate position, this step should correct for those errors. On
average, Goal B would be achieved although results for individual galaxies could still deviate due to significant fiber mis-alignment.

For Data Releases 6 and 7 (DR6 and DR7) of SDSS-I and -II, the flux calibration method used was the same as that
described here. The only difference is that the fibers were 3\arcsec\ in diameter and standard stars were targeted with
16 fibers per plate. The resulting relative spectrophotometric calibration in SDSS-I and -II has an RMS error of 5\% in relative calibration (measured with $g-r$ color) and an RMS error of 4\% in absolute calibration (r magnitude) \citep{SDSSDR6}. In SDSS-III/BOSS, with smaller fibers sizes (2\arcsec), the error was somewhat worse with an 6.3\% RMS error in $g-r$ and 5.8\% RMS error in $r$ for galaxies and stars\footnote{The spectrophotometry error is different for quasar targets in SDSS-III/BOSS, see \cite{Margala15}.} \citep{Dawson13}. 

From a practical standpoint, this is nearly the best approach one could take without significantly greater effort given
the difficulty to determine exactly how the fibers are positioned relative to each star and each science target.
Without that information, it is impossible to separate throughput losses from aperture-induced flux errors. However,
for integral field spectroscopy, this challenge must be overcome.

\subsection{Calibration for integral-field spectroscopy (SDSS-IV/MaNGA)} 
%\section{Principles of IFU spectrophotometry}%what's to correct in spectrophotometry
The goal of integral field spectroscopy is to probe the spatially-resolved information in an extended source. No assumption about the uniformity of any properties of the target would be appropriate. Each aperture element in an IFS instrument (a fiber in a bundle, or a lenslet in a lenslet array) yields a sampling of the seeing-convolved, aperture-convolved surface brightness profile of the target as a function of wavelength.  
In calibrating the flux for each aperture element of an integral field unit (IFU), only the throughput loss should be
corrected, not any aperture-induced flux errors. Therefore, we will separate these two factors using standard star
observations in order to calibrate spectral surface photometry and achieve Goal A.
%determine the throughput loss separately from the aperture-induced flux error in order to calibrate the spectrophotometry for an IFU observation.

Differential atmospheric refraction will still cause each IFU fiber to sample different parts of a target
galaxy at different wavelengths. Rather than attempting to correct for this spatial shift,
we instead simply compute a position array corresponding to the effective location of each IFU fiber on the sky as a
function of wavelength.  When the individual fiber spectra are combined together into a rectified data cube (for
details see Law et al, in prep) the DAR effect will be removed by reconstructing images of the source at each
wavelength using these effective fiber locations.  In other words, our goal here is to correct only for non-geometric
system throughput losses.

If we could measure all of the light from the calibrators (stars) with large, fully-sampled, apertures that delivered the same spectral resolution as our galaxy spectra, then getting the throughput correction would be trivial. This turns out to be difficult in practice. We considered various hardware solutions to separate the throughput loss from the aperture-induced flux error (see Appendix \ref{sec:hardwarechoices}) . Below we first describe what other IFS surveys do for calibration and then present our solution in \S\ref{sec:implementation}. 

%If the IFU has a 100\% fill factor, then there is no aperture loss and getting the throughput correction from the stars would be trivial. However, if the IFU has a fill factor less than 100\%, we would have to find a way to separate the two flux loss factors. 

%Uniqueness of the SDSS observing setup. Plate drilling and plugging system, which allows for simultaneous standard star observation. At the same time, the holes for standard stars have to be drilled beforehand and cannot be moved at the time of the observation. Given the plan of dithered observation, this put constraints on the spectrophotometry methods.

\subsection{Flux calibration in other IFS surveys}

There have been many IFS surveys of galaxies, including SAURON \citep{deZeeuw02}, ATLAS3D \citep{Cappellari11},
DiskMass \citep{Bershady10}, PINGS \citep{Rosales-Ortega10}, CALIFA \citep{Sanchez12}, VENGA \citep{Blanc13}, and SAMI
\citep{Bryant14}. In all of these surveys except SAMI, galaxies are observed one at a time, and standard stars are
observed at different times from the science targets because of instrumental constraints. This practice assumes that
the observing conditions are the same between the science exposures and the calibration exposures, which is not always
true. This is a major difference from the methodology in MaNGA, in which 17 galaxies are observed simultaneously along
with 12 standard stars, enabling independent flux calibration corrections for every exposure. Of particular relevance
in motivating our approach are the DiskMass, PINGS, CALIFA, and VENGA surveys, as all of these make use of fiber bundles with incomplete spatial coverage. 

The PPak instrument used by DiskMass, PINGS, and CALIFA has 2\farcs7 fibers\footnote{Throughout the paper, the fiber
sizes given always refer to the flux-sensitive core of the fibers, not the outer diameter of the buffered fiber.}, slightly larger than MaNGA, so the aperture loss and DAR effects are smaller. The DiskMass Survey did not do flux calibration as the wavelength coverage was very narrow and the main goal of the survey was to constrain kinematics. In the PINGS survey, the throughput correction (which they call nightly sensitivity function) is derived from the standard star observations by applying a monochromatic aperture correction to the standard star spectrum \citep{Rosales-Ortega10}. Alignment offset, wavelength-dependent seeing, and DAR would cause the actual aperture-correction to be wavelength dependent. This was not taken into account.
%a monochromatic aperture covering factor is estimated using simulations and applied to the data at all wavelengths \citep{Rosales-Ortega10}. For observations obtained at relatively high airmass ($>1.2$), this could be problematic. Some observations in PINGS are done at airmass as high as 1.4. 
According to \cite{Rosales-Ortega10}, when normalized at 4861 \AA, the resulting relative calibration has a min-to-max variation of $\pm15\%$ at 3700 \AA\ and $\pm10\%$ at 6850 \AA. %{\bf For CALIFA, the spectrophotometric calibration method has been evolving with time.
For CALIFA, the original spectrophotometric calibration procedure adopted was in essence very
similar to that adopted by PINGS and also did not include the wavelength-dependent aperture correction for standard stars. 
%When normalized at 4600 \AA, the relative calibration has an RMS of $2-3\%$ redward of 3850 \AA, and increases towards the blue end reaching 8\% at 3700-3750 \AA. 
Since late 2013 and for CALIFA data released in DR2 \citep{GarciaBenito15}, an improved calibration scheme was adopted. It uses a set of elliptical galaxies as the calibrator, rather than using standard stars. Because outer regions of elliptical galaxies have very smooth surface brightness profiles, slight alignment offset and DAR would have much less impact on the shape of the spectra. These elliptical galaxies were previously calibrated to the standard spectrophotometric stars by observing both with the PMAS Lens-Array (LArr). As PMAS Lens-Array has a 100\% fill factor, it does not suffer from wavelength-dependent aperture loss. When compared to SDSS images, CALIFA DR2 data have a 5\% RMS calibration error in $g$-band and 6\% in $r$-band. The $g-r$ color has a 3\% RMS error relative to SDSS images. CALIFA applies a final absolute calibration by registering to SDSS broadband images. This step can take out any remaining absolute calibration error in one band but will keep the relative calibration error between different wavelengths.
%The absolute calibration averaged among 200 galaxies, measured by comapring to best-fit stellar population models, is better than 3\% over most of the wavelength and reach 3-5\% around $3700-3850{\rm \AA}$.
%Consequently, the PINGS and CALIFA surveys also make use of broadband photometry of target galaxies to obtain an absolute calibration. 

The VENGA survey \citep{Blanc13} used the VIRUS-P instrument which has 4\arcsec\ fibers. Any wavelength dependence of
the aperture correction for these ``fat'' fibers is probably small enough to neglect provided the observations are done
at reasonably high altitude. Spectroscopic standards were observed with multiple dither positions. The fluxes in
multiple fibers in all dither positions are used to fit a fiber-convolved PSF profile. Then a monochromatic aperture
correction is derived and used to correct the spectra before comparing it with the standard spectrum. The relative
calibration accuracy is estimated to be $\sim8\%$. Afterwards, the absolute calibration is obtained by comparing the synthesized images from the data cube with the broadband optical images.

%\cite{Bershady05} has demonstrated the use flux ratios among fibers to constrain the aperture correction of a central fiber for a single wavelength. 
The SAMI survey employs a two-step flux calibration process \citep{Sharp15,Allen15}. First, a primary
spectrophotometric standard star is observed during the same night as the galaxy observations (but not simultaneously)
to provide a low-order calibration for the wavelength-dependent throughput correction. This is done by fitting a PSF model to the fluxes of multiple fibers in the bundle yielding a wavelength-dependent aperture correction. This is then taken into account in deriving the throughput.  Each plate also includes a secondary standard star, observed simultaneously as the galaxy observations, which provides the telluric correction and an absolute wavelength-independent flux scaling. 
%The stars are all observed with fiber bundles and the fluxes of multiple fibers are used to constrain the PSF of the star and derive the aperture loss as a function of wavelength. 
Comparing the resulting stellar spectra of the secondary standard stars to broadband photometry, the relative calibration in $g-r$ color is 4.3\% with a systematic offset of 4.1\%. By comparing broadband photometry of target galaxies with those obtained from the datacube, the absolute calibration is found to have a systematic offset of 4.4\% and a $1\sigma$ scatter of 28\%. 
As we detail in later sections, the method we adopted for MaNGA is similar to SAMI, but we observe multiple standard stars through fiber bundles simultaneously with the science targets and we use the guider images to facilitate the PSF fitting. 

MaNGA also has much wider wavelength coverage than all of the above surveys. Therefore, the DAR and
wavelength-dependent aperture correction have a more significant impact on the MaNGA data.

\section{Science Requirements for Spectrophotometry}\label{sec:requirements}

The required accuracy of spectrophotometry is determined by the science requirements of the survey. Those that make use
of emission-lines are most sensitive to relative spectrophotometry. One goal is to measure the gas phase metallicity in
star-forming galaxies, which requires the measurement of multiple emission lines including at least \oiiw, \hb, \oiiiw,
\hal, and \niiw. They are spread across nearly 3000\AA\ in wavelength. Thus relative spectrophotometry is crucial. Dust
extinction corrections are also needed in computing some of the indicators that involve widely separated lines, such as \nii/\oii. Extinction estimates are derived from the Balmer decrement, \hal/\hb, which is also sensitive to relative spectrophotometric calibration. 

%MaNGA's specific requirement is to measure gas phase metallicity to 0.1 dex precision when \hb\ is detected at more than 7$\sigma$. This translates to specific requirements on spectrophotometric calibration.
For MaNGA, we require that the uncertainty on spectrophotometric calibration does not dominate the uncertainties on the derived star formation rate and gas metallicities. 

\subsection{Calibration requirement on \hal\ and \hb}

First, we describe how the error on spectrophotometry could translate to the error on extinction and star formation rate (SFR). 

Below, we use $C_\lambda$ to denote the flux calibration vector which needs to be multiplied with the raw flux to get the calibrated flux. 
For line fluxes, we use $F_{r}(\hal)$ to denote the raw fluxes measured before applying the spectrophotometric calibration, $F_o(\hal)$ to denote the flux after calibration, and $F_c(\hal)$ to denote the flux after extinction correction\footnote{Here, we are referring to extinction intrinsic to the source, not the foreground extinction in the Milky Way}. Given these definitions, we have
\begin{equation}
F_o(\hal) = F_r(\hal) C_{\rm H\alpha} 
\end{equation}
The uncertainty on $F_o(\hal)$ should follow
\begin{equation}
\left(\sigma_{F_o(\rm H\alpha)} \over F_o(\rm H\alpha)\right)^2 = \left(\sigma_{F_r(\rm H\alpha)} \over F_r(\rm H\alpha)\right)^2 + \left(\sigma_{C_{\rm H\alpha}} \over C_{\rm H\alpha}\right)^2
\label{eqn:sigma_F_o}
\end{equation}
We define $c_1 = C_{\rm H\alpha} /C_{\rm H\beta}$, the relative calibration between \hal\ and \hb.  
Extinction is usually derived using the Balmer decrement. We define $r=F_o(\hal)/F_o(\hb)$ and let $\sigma_r$ denote the uncertainty of $r$ such that
\begin{equation}
r = {F_o(\hal) \over F_o(\hb)} = {F_r(\hal) \over F_r(\hb)}c_1
\end{equation}
\begin{equation}
\left(\sigma_r \over r\right)^2= \left({\sigma_{F_r(\rm H\alpha)} \over F_r(\hal)}\right)^2 + \left({\sigma_{F_r({\rm H\beta})} \over F_r(\hb)}\right)^2 + \left(\sigma_{c_1} \over c_1\right)^2
\label{eqn:sigma_r}
\end{equation}

Taking the Case B Balmer decrement of 2.863 at $T= 10^4{\rm K}$ and $n=10^2{\rm cm}^{-3}$ \citep{OsterbrockBook}, we have  
\begin{equation}
E(B-V) = {2.5\over (k_{\rm H\beta} - k_{\rm H\alpha}) }\log{r\over2.863}
\end{equation}
Here $k_{\rm H\alpha}$ indicates the total-to-selective extinction for \hal. The uncertainty on $E(B-V)$ is
\begin{equation}
\sigma^2_{E(B-V)} = \left({2.5\over (\ln10) (k_{\rm H\beta} - k_{\rm H\alpha}) }\right)^2 \left( \sigma_r \over r\right)^2
\label{eqn:sigma_EBV}
\end{equation}
The extinction-corrected \hal\ flux is 
\begin{align*}
F_c(\hal) &= F_o(\hal)10^{0.4 A_{\rm H\alpha}} \\
 &= F_o(\hal)10^{0.4 k_{\rm H\alpha} E(B-V) }
\end{align*}
The uncertainty on $F_c$ can be derived as
\begin{equation}                      
\left(\sigma_{F_c(\rm H\alpha)} \over F_c(\hal) \right)^2= \left(\sigma_{F_o(\rm H\alpha)} \over F_o(\hal)\right)^2 + (0.4 k_{\rm H\alpha} \ln10 \sigma_{E(B-V)})^2
\label{eqn:sigma_Lc}
\end{equation}

%$L_c$ is related with $L_o$ through 
%\begin{align*}
%L_c(\hal) &= L_o(\hal) 10^{0.4 A_{\rm H\alpha}} \\
%                      &= L_o(\hal) 10^{0.4 k_{\rm H\alpha} E(B-V)} 
%\end{align*}
%The uncertainty on $L_c$ can be derived as
%\begin{equation}                      
%\left(\sigma_{L_c(\rm H\alpha)} \over L_c(\hal) \right)^2= \left(\sigma_{L_o(\rm H\alpha)} \over L_o(\hal)\right)^2 + (0.4 k_{\rm H\alpha} \ln10 \sigma_{E(B-V)})^2
%\label{eqn:sigma_Lc}
%\end{equation}
%Assuming zero uncertainty on distances, and using Equation~\ref{eqn:sigma_EBV}, we have 
%\begin{equation}
%\left(\sigma_{L_c(\rm H\alpha)} \over L_c(\rm H\alpha)\right)  = \left(\sigma_{F_o(\rm H\alpha)} \over F_o(\hal)\right)^2 + \left(k_{\rm H\alpha} \over k_{\rm H\beta} - k_{\rm H\alpha}\right)^2 \left(\sigma_r \over r\right)^2
%\end{equation}
A common estimate of star formation rate is derived from the extinction-corrected \hal\ luminosity ($L_c(\hal)$). 
Adopting the SFR calibration given by \cite{Kennicutt98}, ${\rm SFR (M_\odot~year^{-1})} = 7.9\times10^{-42} L_c(\hal) (\rm ergs~s^{-1})$, and assuming zero uncertainty on distances, we have 
\begin{equation}
\left(\sigma_{\rm SFR} \over {\rm SFR} \right)^2 = \left(\sigma_{L_c(\rm H\alpha)} \over L_c(\hal)\right)^2 = \left(\sigma_{F_c(\rm H\alpha)} \over F_c(\hal)\right)^2
\label{eqn:sigma_SFR_L_F}
\end{equation}

Combining Equations (\ref{eqn:sigma_F_o}), (\ref{eqn:sigma_EBV}), (\ref{eqn:sigma_Lc}) and (\ref{eqn:sigma_SFR_L_F}), we have
\begin{align} 
\left(\sigma_{\rm SFR} \over {\rm SFR} \right)^2 =&\left(\sigma_{F_c(\rm H\alpha)} \over F_c(\hal)\right)^2 \\
=& \left({\sigma_{F_o(\rm H\alpha)} \over F_o(\rm H\alpha)}\right)^2 + \left(k_{\rm H\alpha} \over k_{\rm H\beta} - k_{\rm H\alpha}\right)^2 \left(\sigma_r \over r\right)^2 \\ \nonumber
=&\left(\sigma_{F_r(\rm H\alpha)} \over F_r({\rm H\alpha})\right)^2 +\left(\sigma_{C_{\rm H\alpha}} \over C_{\rm H\alpha}\right)^2 \\ 
 &+ \left(k_{\rm H\alpha} \over k_{\rm H\beta} - k_{\rm H\alpha}\right)^2 \left(\sigma_r \over r\right)^2 \label{eqn:sigma_SFR}
\end{align}

Adopting the dust attenuation law of \cite{O'Donnell94} and $R_{\rm v}=3.1$, we have $k_{\rm H\alpha} = 2.519$ and $k_{\rm H\beta}=3.663$. Combining equations (\ref{eqn:sigma_SFR}) and (\ref{eqn:sigma_r}), we have
\begin{align}
\left(\sigma_{\rm SFR} \over {\rm SFR} \right)^2&= 5.85\left(\sigma_{F_r(\rm H\alpha)} \over F_r({\rm H\alpha})\right)^2 + 4.85 \left(\sigma_{F_r(\rm H\beta)} \over F_r({\rm H\beta})\right)^2  \nonumber \\
&+4.85 \left(\sigma_{c_1} \over c_1\right)^2 + \left(\sigma_{C_{\rm H\alpha}} \over C_{\rm H\alpha}\right)^2
\label{eqn:finalhahb}
\end{align} 

The right hand side of the Equation~\ref{eqn:finalhahb} contains 4 terms. The first two terms are related with the fractional errors of the raw measurements of \hal\ and \hb, and the latter two are related with the fractional errors of the relative flux calibration and the absolute calibration. The calibration errors would be uniform across each galaxy, but the errors of \hal\ and \hb\ would depend on the strength of the lines. One of our science requirements is to measure SFR surface density to 0.2 dex. Flux calibration would not dominate the total error when emission lines are weak, but it would dominate when emission lines are strong. We therefore require that, in regions of strong line detections, the error on SFR estimates due to flux calibration {\it alone} needs to be better than 0.05 dex (a fractional error of 11.5\% on SFR estimates). This would ensure that the calibration error be subdominant anywhere \hb\ is not measured to better than 19$\sigma$ (5.2\% fractional error; $0.115^2 = 4.85\times0.052^2$). We split this error budget (0.05 dex on SFR) equally between the relative calibration and the absolute calibration --- 3rd and 4th term in Equation~(\ref{eqn:finalhahb}). This means that the relative flux calibration between \hal\ and \hb\ needs to be measured to better than 3.7\%, and the absolute calibration around \hal\ needs to be better than 8.1\%.

\subsection{Requirement on relative calibration between \nii\ and \oii}
Similarly, we can derive how gas-phase metallicity measurements are affected by the spectrophotometry error. For example, one important gas phase metallicity indicator is the \niibw/\oiiw\ ratio \citep{KewleyD02}. It is one of the best indicators but requires good spectrophotometric calibration and good extinction corrections. Here we denote the ratio in flux calibration correction between \nii\ and \oii\ as $c_2$, the raw flux measurements as $F_{\rm line}$, and the extinction-corrected \nii/\oii\ ratio as $R$. $R$ can be expressed as 
\begin{align}
%\begin{equation}
R_{\rm [NII]/[OII]} &= {F_{\rm [N~II]} 10^{0.4 A_{\rm [N~II]}}\over F_{\rm [O~II]} 10^{0.4 A_{\rm [O~II]}} } c_2 \nonumber\\
 &= {F_{\rm [N~II]}\over F_{\rm [O~II]}} 10^{0.4(k_{\rm [N II]} - k_{\rm [O II]})E(B-V)} c_2
\end{align}
Combining with Equations~\ref{eqn:sigma_r} and ~\ref{eqn:sigma_EBV}, the fractional error on $R$ can be written as
\begin{align}
%\begin{equation}
\left(\sigma_R \over R \right)^2 =& {\sigma^2_{F_{\rm [N~II]}} \over F^2_{\rm [N~II]}} + {\sigma^2_{F_{\rm [O~II]}} \over F^2_{\rm [O~II]}} + {\sigma^2_{c_2} \over c_2^2} \nonumber \\ 
&+ \left(k_{\rm [N~II]} - k_{\rm [O~II]} \over k_{\rm H\beta} - k_{\rm H\alpha}\right)^2 \left({\sigma^2_{F_{\rm H\alpha}} \over F^2_{\rm H\alpha}} + {\sigma^2_{F_{\rm H\beta}} \over F^2_{\rm H\beta}} + {\sigma^2_{c_1} \over c^2_1} \right) \nonumber \\
=& {\sigma^2_{F_{\rm [N~II]}} \over F^2_{\rm [N~II]}} + {\sigma^2_{F_{\rm [O~II]}} \over F^2_{\rm [O~II]}} + 3.57 \left({\sigma^2_{F_{\rm H\alpha}} \over F^2_{\rm H\alpha}} + {\sigma^2_{F_{\rm H\beta}} \over F^2_{\rm H\beta}}\right) \nonumber\\
&+ {\sigma^2_{c_2} \over c_2^2} + 3.57 {\sigma^2_{c_1} \over c^2_1}
.
\label{eqn:sigma_n2o2}
%\end{equation}
\end{align}
In the above equation, we have adopted the extinction law given by \cite{O'Donnell94} to compute the coefficient involving the $k$ factors. 

According to \cite{KewleyD02}, \nii/\oii\ is a good metallicity indicator for regimes where $\log({\rm O/H})+12$ is greater than 8.6 (approximately $\log(\nii/\oii) > -1$). In this regime, 
\begin{align}
\log ({\rm O/H}) & + 12 = \nonumber\\
 \log [1.540 &+ 1.266 \log R + 0.168 \log^2 R ] + 8.93 {\rm .}
\end{align}
The error on metallicity would be
\begin{align}
\sigma(\log ({\rm O/H} & )) =  \nonumber\\
{1\over \ln^2 10} &{1.266+ 2\times 0.168 \log R \over 1.540+1.266 \log R + 0.168 \log^2 R } {\sigma_R \over R}
\end{align}

One of our science requirements is to measure gas phase metallicity (O/H) to 0.1 dex. Fix $\sigma(\log ({\rm O/H}))$ to 0.1, we can derive the allowed maximum fractional error on R as a function of R. The smaller R is, the tigher the constraint is. At R=0.1 (corresponding to $\log({\rm O/H})+12 = 8.57$), the fractional error on $R$ needs to be smaller than 25.2\% to meet the requirement. Again, the flux calibration will only dominate the error when emission lines are strongly detected. We require that, in regions of strong line detections, the fractional error on $R$ due to flux calibration {\it alone} to be less than 10\%, which would translate to a maximum error of 0.04 dex on O/H. There are two terms in Equation~\ref{eqn:sigma_n2o2} that are related with flux calibration. Splitting the error budget equally between these two terms, we result at the requirements on the fractional error of $c_1$ and $c_2$: $c_1$ needs to be measured to better than 3.7\%, and $c_2$ needs to be measured to better than 7\%. These ensure that the flux calibration error be subdominant until \hb\ is measured to better than $19\sigma$. 

%Again, the biggest contributor to the total error budget is likely to be uncertainties on \hb,
%through the use of \hb\ flux in the extinction correction. For flux calibration to not dominate the error, we require the fractional error contributed by flux calibration to be a factor of 3 smaller than the contribution of \hb\ error (a factor of 9 smaller when squared) when \hb\ is measured to 10\%. Thus, the error on the ratio of flux calibration correction between \nii\ and \oii\ needs to be measured to better than 6.3\%.
%
\subsection{Requirement on uniformity of calibration among exposures}
In addition, we also have a requirement on the uniformity of the flux calibration from exposure to exposure. MaNGA combines multiple exposures taken at 3 different dither positions to synthesize a filled data cube. We need the input exposures to have consistent flux calibration. As simulated by \cite{Law15}, 
if each exposure has a dither-dependent systematic flux calibration error with an RMS of 5\%, a 1\% pixel-to-pixel
error in the reconstructed data cube would result.  If the flux calibration errors are uncorrelated with dither position then they will average out across many exposures and not present a strong requirement on flux calibration accuracy.
%if exposures at the three different dither positions have a systematic flux calibration error with an RMS of 5\% but fixed for each dither, it would lead to a 1\% pixel-to-pixel variation in the reconstructed data cube. If the flux calibration error is random among the dither position, then they average out and we could tolerate much worse calibration errors. 

\bigskip
To summarize, we require the relative flux calibration between \hal\ and \hb\ to be measured to better than 3.7\%, and
that between \nii\ and \oii\ to be measured to better than 7.0\%. Given our conservative requirements, even if the
calibration accuracy is worse by a factor of 3, it would still contribute to the error budget on SFR and metallicity
measurements as would a 16\% error on \hb. We also require the systematic calibration difference among exposures to
have an RMS less than 5\% for the majority of the wavelength range including all the emission lines mentioned here. 

\section{Flux Calibration Method in MaNGA} \label{sec:implementation}% detailed implementation of the calibration method

\subsection{Mini-bundles}
MaNGA adopts hexagonally packed fiber bundles with 2\arcsec\ fiber core diameter and 2.5\arcsec\ center-to-center spacing. This yields a fill factor of 56\%. To approach critical sampling of the focal plane PSF, we carry out dithered observations and obtain a uniform reconstructed effective PSF in the stacked data cubes. (For discussion of the observing strategy, see \citealt{Law15}.) 
While the fibers do not provide 100\% coverage for any given exposure, following our approach described in \S\ref{sec:principle} (Goal A), we do not try to correct for the flux falling into the gaps between fibers, as it is unknown. The knowledge about those missing regions can only be reconstructed through combining dithered observations. 

As discussed in \S\ref{sec:principle}, we need to separate the throughput loss factor from any aperture-induced flux
error. We also have to calibrate each exposure individually as the atmospheric transparency can change with time. These
require that we either obtain all the flux included in the PSF of observed standard stars or figure out a way to
measure the fraction of the PSF sampled by standard star fiber apertures as a function of wavelength. 

The original methods used by SDSS-I to -III will not work well for MaNGA because the miscentering of the fiber relative to the stellar calibration source will be unknown due to drilling error, the clearance between ferrule and the plate hole, the uncertainty in the proper motion of the star, and field differential refraction. 
%Even if we assume that we could guide the telescope perfectly and could model the differential atmospheric refraction, the AR-induced quadrupole distortion of the field could also lead to significant misalignment between the fiber aperture and the target.
%we cannot correct for the quadrupole term, and the subsequent, cumulative miscentering will bring large uncertainty in the exact aperture covered by the fiber. 
Therefore, we cannot accurately predict what fraction of the PSF flux is falling into the fiber at each wavelength even though the PSF can be obtained from the guider camera (\S\ref{sec:guider}).

%Therefore, spectrophotometry calibration for an IFU instrument needs to separate the throughput loss from the aperture loss. 
%For MaNGA, we also have an added complication, because we are going to conduct dithered observations. 

%The final method we investigated is to use the 7-fiber hexagonal bundle described above directly to target standard stars. 
Our solution is to target standard stars with 7-fiber hexagonal bundles with the same fiber size and fill factor as the science bundles. Given the gaps between the fibers, not all the light will be collected. However, the relative flux ratios between the 7 fibers allow an accurate determination of the actual position of the star image inside the bundle. Given a priori knowledge about the PSF shape of the star obtained from the guider images and theoretical knowledge about differential atmosphere refraction, we can accurately reconstruct the fraction of light falling into each fiber at each wavelength. This allows us to estimate the aperture loss separately from the throughput loss. %A similar technique was applied in the VENGA survey \citep{Blanc09} 

In the final MaNGA instrument configuration, we use 12 of these 7-fiber mini-bundles per cartridge\footnote{A cartridge is a large thick disk containing the fiber assembly, the pseduo-slits formed by the fibers, and the structure to hold the plate to be observed. Multiple cartridges are prepared each night for efficient observations. Each cartridge is installed with a plate during the day and fibers in that cartridge are plugged into the plate. At night, when changing field of observation, observers remove the previous cartridge on the telescope and install the cartridge containing the next plate. The cartridge changing process takes just several minutes.} 
to target 12 standard stars. They feed two spectrographs with 6 mini-bundles per spectrograph, which are further grouped into two fiber assemblies with 3 mini-bundles each. Details of these and how they are organized on the slithead can be found in \cite{Drory15}.

\subsection{Measuring the PSF with the Guider} \label{sec:guider}
The guider system is important in this process as it provides first-order knowledge about the size and shape of the PSF. We briefly describe the guider system here. Guiding is achieved using 16 coherent imaging fiber bundles plugged in the same plate as all the science fibers. These coherent fiber bundles are collections of thousands individual fiberoptic strands assembled together so that the relative orientation of the individual strands is maintained throughout the length of the bundle. Most of the guide bundles used here are $450 \mu m$ in diameter and each contain $\sim10,000$ picture elements with each element being only a few microns across. This is not to be confused with the large fiber bundles for science, which has 19-127 fibers with the core of each fiber having a diameter of $120\mu m$ and manufactured in completely different manners. 

These coherent imaging bundles are routed to a guider block on the side of the cartridge and imaged by a guide camera.  
Among the 16 guide bundles, 2 of them are 24\arcsec\ in diameter and they are used for field acquisition. The other 14 guide bundles are 7\arcsec\ in diameter and are used for guiding. Among these 14, 4 are positioned with their surface 400 ${\rm \mu m}$ above the plate surface, 4 are positioned 400 ${\rm \mu m}$ below the plate surface, and 6 are positioned at the plate surface as are the science fibers. This design helps focus the telescope via a comparison of the PSF obtained with the guide bundles above, at, and below the plate surface. The guide bundles are distributed across the plate and provide an estimate for the scale of the field, since the guider images reveal how the stars are offset from their expected positions. The scale of the field can be adjusted by tuning the distance between the primary mirror and the secondary mirror.

During observations, the guider system determines the optimal axis, rotation, and scale adjustments to the telescope that would minimize the distances of the 14 stars from the image centers of their respective guide bundles (determined by the flat image). Under typical seeing conditions, the guider takes an exposure every 27 second (exposures are 15 seconds with 12 seconds overhead). There are roughly 33 frames per 15 minute science exposure. We stack the guider images together to obtain an effective PSF for the science exposure. This is a time-average of the varying seeing during the exposure and also includes the effect of guiding uncertainties. 

We fit each guide star with a double Gaussian with freely varying amplitudes and widths. We choose double Gaussians to model the PSF because they provide a sufficient approximation to the actual PSF within a diameter of $\sim4\times{\rm FWHM}$ and are very fast to compute. The 8\ in-focus guide stars give PSFs that sometimes can vary by as much as 0.1-0.2\arcsec\ in FWHM. The source of this variation is not completely understood, but a few potential causes have been identified. 

First, the curvature of the plate does not conform perfectly to the designed focal plane shape with an error up to a 100 microns. Given the f/5 focal ratio of the telescope and the plate scale of $60.455{\rm \mu m/arcsec}$, focal plane mismatch should contribute at most a 0.33\arcsec-diameter broadening kernel to the PSF. For 1.5\arcsec\ seeing (FWHM), the combined PSF should only be broadened by 0.03\arcsec in FWHM. Therefore, it cannot explain the differences.

Second, the guider output block could have a slight tilt relative to the optical axis of the guider camera. Different guider probes are also not perfectly coplanar to each other, with typical offsets expected to be less than $25{\rm\mu m}$. The guider camera has a much faster focal ratio of f/1.4. Thus, these coplanar offsets are more likely the culprit. If this is the case, it would imply that the PSF seen by the guider and the science IFUs are much more uniform than what the guider camera implies.

Therefore, we pick the sharpest PSF among the six in-focus guide stars as the reference PSF. We denote this PSF profile as $p_0(r,\theta)$. In our case, we model this as a circularly symmetric profile so there is no angular dependence but we keep $\theta$ in the formula to indicate it is a 2D profile. This is modeled from guider images at an effective guiding wavelength of 5400 \AA.

\subsection{Predicting the wavelength-dependent PSF}
%Modifying the PSF model according to the optics}

Next, we need to use the measured PSF at the guide wavelength to predict the wavelength-dependent PSF at the position of the standard star target. There are several factors affecting the PSF at different wavelengths. First, the seeing varies as $\lambda^{-1/5}$ \citep{Fried66, Boyd78}. We scale the PSF accordingly depending on its wavelength. 
%Fried 1966a, JOSA, 56, 1372; 
%Fried 1966b, JOSA, 56, 1380; 
%Boyd, 1978, JOSA,68, 877; JOSA = Journal of the Optical Society of America
\begin{equation}
p_\lambda (r) = p_0(r (\lambda/\lambda_0)^{1/5})
\end{equation}
Here, $\lambda_0$ is the guiding wavelength and is equal to 5400 \AA.

\begin{figure}
\begin{center}
\includegraphics[totalheight=0.15\textheight,viewport=0 0 500 500, clip]{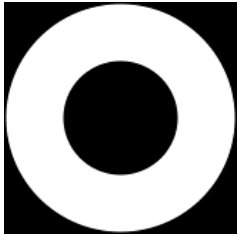}
\vspace*{0cm}
\caption{\small The convolution kernel used in modeling the effect of a focus offset on the PSF.}
\label{fluxcal:fig:focuskernel}
\end{center}
\end{figure}

Second, the telescope focal plane changes with wavelength resulting in focus offsets as a function of wavelength and position on the plate.The focal plane shapes as designed are given by \cite{Gunn06} in Table 5 of that paper. We interpolate to obtain the focus offset at each wavelength according to target position on the
plate.  The out-of-focus PSF should be computed by convolving the in-focus PSF with a ring kernel as shown in
Figure~\ref{fluxcal:fig:focuskernel}, which is the telescope pupil. The outer diameter of the ring is set to 1/5 of the focus offset because the telescope has an f/5
beam. The inner diameter of the ring equals 1/10 of the focus offset, which is set by the size of the secondary mirror. We convert the sizes in length units to angular units using the plate scale of $3.62730{\rm mm/arcmin}$ \citep{Gunn06}. 
We denote this kernel $k_{\lambda, d}(r,\theta)$, where $d$ is the distance from the center of the plate.

The PSF we would observe is then 
\begin{equation}
\phi_{\lambda,d}(r, \theta) = p_\lambda(r,\theta) \ast k_{\lambda,d}(r)
\end{equation}

This
yields the PSF expected as a function of wavelength and position on the plate.

For computational convenience, we also convolve the above PSF model with the fiber aperture (a 2\arcsec\ diameter circular step function) to obtain the final profile from which we can simply interpolate to obtain the flux one would get in any 2\arcsec\ fiber in the 7-fiber mini-bundle. 
%In essence, imaging spectroscopy is just a sampling of the seeing-convolved, fiber-convolved surface brightness profile at each wavelength. 

\subsection{Typing the star}

%The standard stars used by SDSS are dwarf F stars. %The exact stellar type of each star is determined from its observed spectrum by fitting theoretical models to a normalized spectrum. 
The standard stars we select for MaNGA are late-F type main-sequence stars. We require the stars to have no nearby bright neighbors to ensure the wings of the PSF is not contaminated. We select stars with observed magnitudes between 14.5 and 17.2\ in the $g$-band. If we cannot find enough stars for a field, we move the faint limit to 17.7, or 18.2 if necessary. Late-F type main-sequence stars have an absolute magnitude ranging roughly between 2.5 and 4\ in $g$-band. Our magnitude range ensures that they are at least 1 kpc away. For the MaNGA galaxy program fields, which are all at galactic latitude ($b$) higher than $20^\circ$, these stars are certainly halo stars and are beyond most of the galactic dust.

%Therefore, we need to apply the reddening to the model. 
%The final flux calibrated spectra are also calibrated to that as observed outside the galactic plane. 

%This part of the code is similar to the procedure in the BOSS pipeline. 
%but we combine spectra from all three dither positions for typing to maximize signal-to-noise of the observed spectrum. 
The reduction pipeline provides a sky-subtracted spectrum for each fiber for each exposure. We first divide these spectra by an initial estimate of the throughput vector, which is the average throughput derived from tens of plates processed by an earlier run of the pipeline. For each mini-bundle, the fiber with the maximum total flux over the whole wavelength range is selected as the reference fiber whose spectrum is used to determine the model spectrum. In fitting the theoretical models, we adopt the same algorithm as used in the SDSS Legacy and SDSS-III/BOSS pipelines, as described in \S\ref{sec:oldmethod}. The resulting model spectra are scaled to match the r-band PSF magnitude of the stars.

%a running-median-filtered overall continuum is divided from both the data and the model so that the unknown flux calibration does not affect the typing of the star. We use a grid of Kurucz models with a range of temperature, [Fe/H], and surface gravity, same as that used in the SDSS Legacy and SDSS-III BOSS pipelines. The best fit model is first reddened according to \cite{SchlegelFD98} dust map and \cite{O'Donnell94} extinction law, and then scaled according to the r-band magnitude of the star. 

\subsection{Fitting for the flux ratios among fibers in a mini-bundle}
\label{fluxcal:sec:analysis}

%In the test run, we had 8 IFU bundles with sizes between 19 and 127 fibers. We use the 7 central fibers of each bundle to mimic the mini-bundle.

To accurately model the aperture loss of a single fiber in the mini-bundle, we first need to know the position of the star relative to the bundle and the PSF. The exact position of the star is uncertain due to uncertainty in astrometry and proper motion, drilling errors, the positional uncertainty of ferrule in its hole, and telescope pointing error. The exact PSF seen by each bundle could also differ from what the best guide star sees due to two reasons: (a) the plate shape is not perfectly matching the focal plane, (b) the smearing of the standard star during the science integration from guiding could be different from the smearing of the guide star, due to the constant scale and rotation adjustments applied by the guider feedback loop which are imperfect. Therefore, we use the flux ratios among the 7 central fibers as a function of wavelength to constrain the position of the star and the size of the PSF.

\begin{figure*}
\begin{center}
\includegraphics[width=1.0\textwidth, viewport = 20 50 760 570, clip]{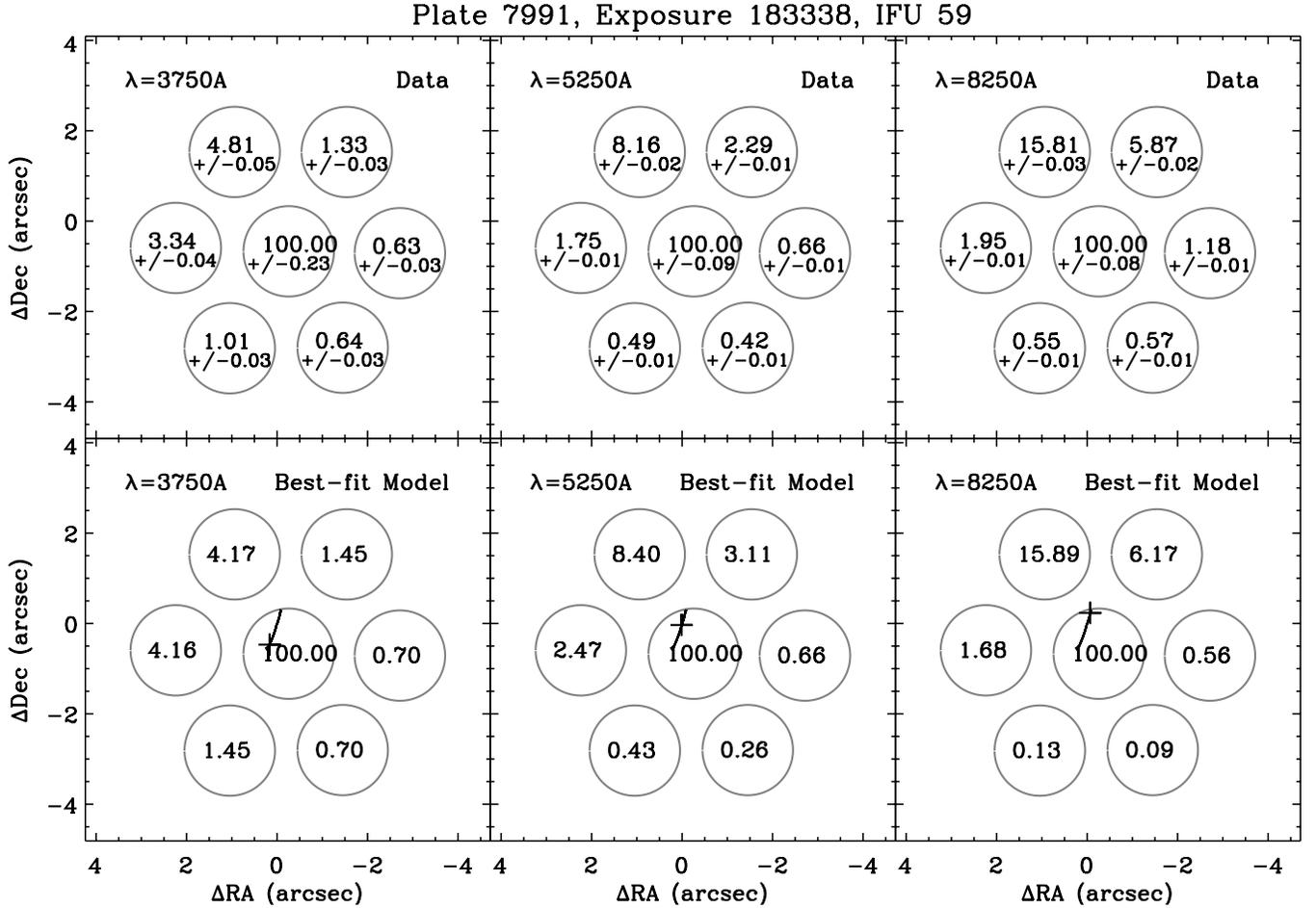}
\vspace*{0cm}
\caption{\small This figure illustrates how we model the flux ratios among fibers at different wavelengths to constrain the position of the star relative to the bundle. The three columns show three different wavelengths (left: 3750 \AA, middle: 5250 \AA, right: 8250 \AA). The top panels show the flux ratio (in percentage) of each fiber relative to the fiber with the most flux. The bottom panels show the ratios in the best fit model. The slanted line in the bottom panels indicate the constrained position of the star and the DAR vector. The `+' symbol indicates the position of the star at the plotted wavelength. Note the relative flux in the two top fibers increases with wavelength as the position of the star image moves up. This exposure is observed at an airmass of 1.13\ in the northern sky --- higher Declination corresponds to lower altitude in this case.}
\label{fluxcal:fig:fluxratio}
\end{center}
\end{figure*}

Figure~\ref{fluxcal:fig:fluxratio} illustrates our method. We choose the fiber with the highest total flux within
3500--10500 \AA\ as the reference fiber. We then sum the flux in eight wide wavelength windows (3500-4000 \AA, 4000-4500 \AA, 4500-5000 \AA, 5000-5500 \AA, 5500-6500 \AA, 6500-7500 \AA, 7500-9000 \AA, and 9000-10500 \AA) for all fibers and take the ratio
between each fiber and the reference fiber for each wavelength window. We run a Markov Chain Monte Carlo with four variables: x, y
position of the star, scaling and rotation of the differential atmosphere refraction vector. Given a set of these four
parameters, we compute the expected flux ratios from the PSF model. Taking the difference between the observed ratios and the
model ratios we compute the $\chi^2$ for each step and use the MCMC chain to find the $\chi^2$ minimum. The chain often
converges within a couple hundred steps. We run it for 500 steps and take the solution giving the minimum $\chi^2$. We then
scale the PSF to smaller and larger sizes and find the minimum $\chi^2$ for each PSF size. Fitting the minimum Chi Square as a
function of PSF size by a quadratic function, we find the PSF size that yields the best fit to the flux ratios among fibers, along
with the position of the star and DAR. With this best PSF constrained, we rerun the chain for 2000 steps to find the best solution for offsets, rotation and scale. The reason we do not include PSF size as one variable in the MCMC is that the computation of the PSF is a slow process as it involves two convolution procedures.  

Throughout this process, whenever summing the observed flux within each wavelength window, we weight each pixel by the inverse variance. The same weighting is applied to the model spectrum. Therefore, the data and the best-fit model should have nearly identical effective wavelength in each wide wavelength window.

\subsection{Deriving the throughput loss}

\begin{figure*}
\begin{center}
\includegraphics[angle=180, width=\textwidth,viewport=50 30 800 315, clip]{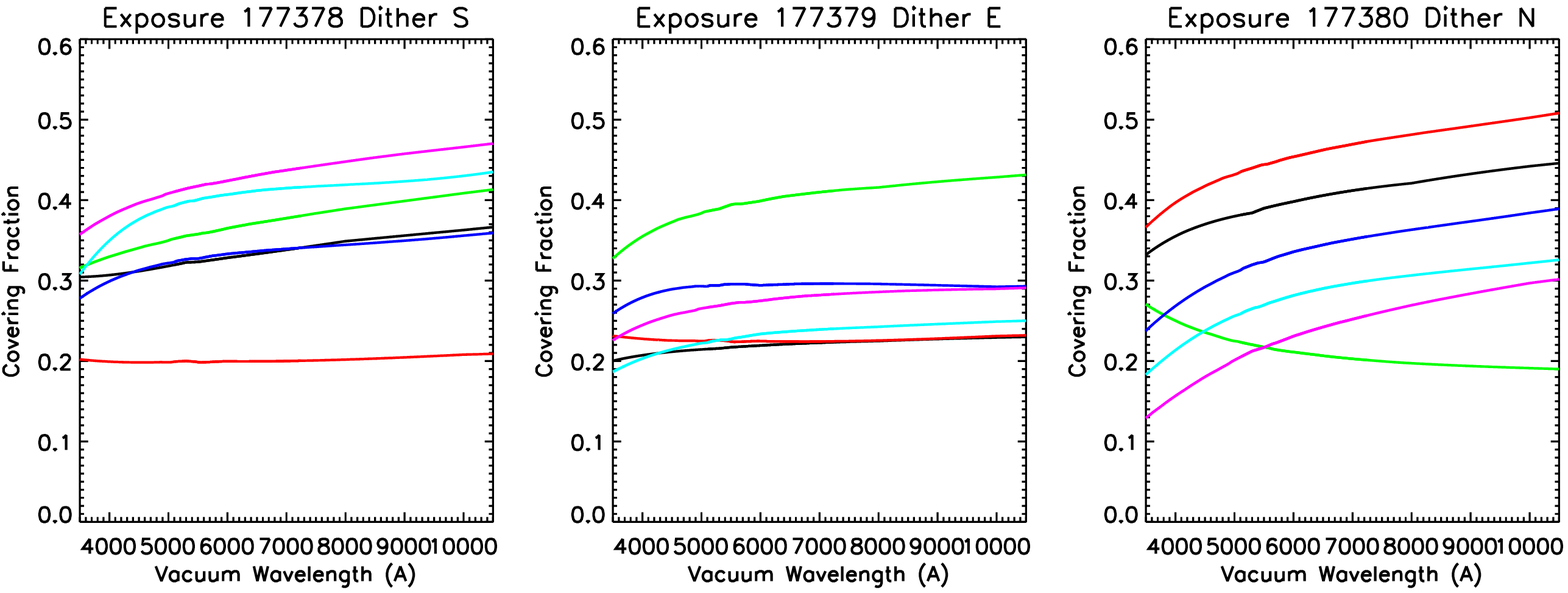}
\includegraphics[angle=180, width=\textwidth,viewport=50 30 800 315, clip]{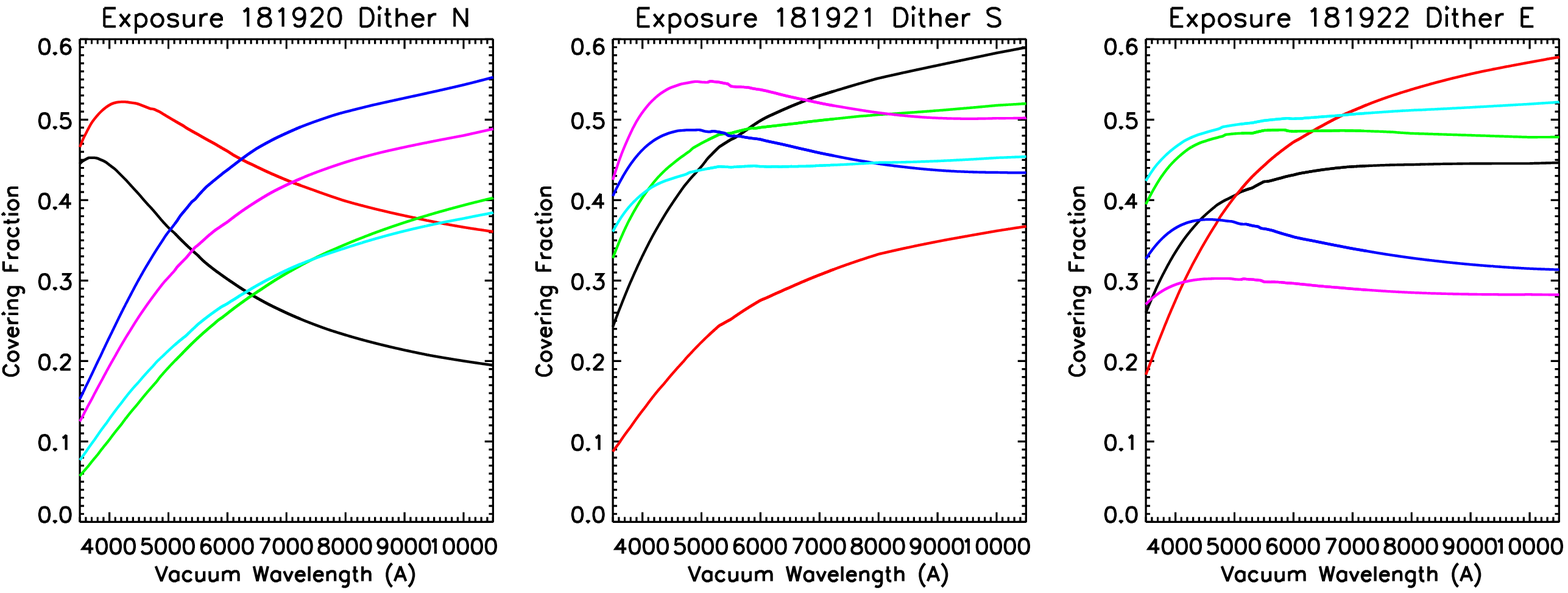}
\caption{The PSF-covering fractions of the reference fiber in six mini-bundles as a function of wavelength are shown for three dithered exposures in a set. The top row shows a low airmass observation and the bottom row shows a high airmass observation. Each color indicates a unique mini-bundle with consistent color-coding between the two rows. The PSF-covering fraction for the same star can differ signficantly between different airmasses.}
\label{fluxcal:fig:covfn}
\end{center}
\end{figure*}

Given the best fit model, we derive the PSF-covering fraction of the fibers, which is defined as the fraction of the flux in a PSF covered by a fiber as a function of wavelength. Figure~\ref{fluxcal:fig:covfn} shows examples of the derived covering fractions for the central fibers in six mini-bundles on one spectrograph for three dithered exposures in a set, and for two airmasses with different levels of atmospheric refraction.
We then compute the expected flux of the star by multiplying the theoretical model spectrum with the PSF-covering fraction. Dividing the observed flux by the expected flux from the theoretical model yields the effective correction vector for each star.

%\caption{\small Flux correction vector (effective throughput) derived from the brightest star among those targeted by IFUs on the low-airmass standard plate.  This curve is different from the product of the throughput of all optical elements, the atmosphere, and the QE of the CCD because it also contains a correction vector that is effectively multiplied in the flat-fielding step, giving it its shape.}

%\subsection{``Averaging'' the correction vector among all stars}

The correction vectors derived from the six standard stars on each spectrograph differ slightly in their normalization and the low-order shape. Possible sources of this variation include error in the magnitude of the star, error in the derived covering fractions, error in the model determination, error in the flux extraction from the 2d spectra, and the variation of throughput due to airmass differences. The covering fraction error appears to be the dominant source. In some cases, we cannot find a satisfactory fit to all the flux ratios to within the measurement uncertainty. This is probably due to the simplicity of our PSF model and neglect of the guiding error. The true PSF is not circular at all positions on the plate and can be more asymmetric at wavelength extremes. The typical guiding stability of the SDSS Telescope is about 0.12\arcsec. Compared to the typical seeing at the site --- 1.5\arcsec, most of the time the smearing caused by guiding error should contribute minimally to the final PSF. However, under superb seeing conditions (1.0\arcsec\ or better) and at high altitude ($>80^\circ$) where guiding is worse for this Alt-Az telescope, the guiding error starts to contribute significantly to the integrated PSF over the 15 minute exposure. These factors could contribute to the inconsistency among the correction vectors derived for the standard stars.

%The model spectra are scaled according to the r-band PSF magnitude of the standard stars. In principle, the derived correction vectors should all agree in normalization at the r-band. However, there is disagreement among them. The cause of the disagreement is not dominated by photometry error because it changes among the different exposures for the same set of stars. It is likely due to the error in the absolute covering fraction we derived. 
%(Note: adding the brightest two or three fibers together would probably lower the error in the total covering fraction, which I will try. Though it may lower the total S/N if the S/N difference is large between the brightest and the 2nd brightest. Further improvements on the fitting procedure (Sec 5.4) would also improve the absolute calibration.)

Among 54 exposures taken during the commissioning run in March 2014 on 4 different plates, the average fractional RMS of the normalization difference among the 6 stars per spectrograph was 6\%, and better than 13.4\% in 95\% of the exposures. The resulting corrections are the mean of all stars and thus have a much smaller uncertainty. Before construting an effective average of the correction vectors, we reject outliers using a series of criteria. Notably, we reject stars that satsify at least one of the following criteria: 
\begin{enumerate}
\item Having a median S/N (among all pixels) lower than 1/3 of the median median-S/N of all stars. 
\item Having a $\chi^2$ from stellar model fit that is more than 3 times larger than the median $\chi^2$ of all stars.
%\item stars whose flux ratio fit has very large $\chi^2$ (higher than 100 or the median $\chi^2$ of all stars, whichever is larger). 
\item Having a $\chi^2$ from flux ratio fit that is higher than 100 or the median $\chi^2$ of all stars, whichever is larger.  
\item We evaluate the median level of each correction vector in two wavelength windows (5300-5350 \AA\ for the blue camera and 7800-8000\AA\ for the red camera). Stars are rejected if their correction vectors are greater than $3\sigma$ (or 10\%, whichever is larger) away from the median levels among all stars in either the blue or the red window. Here, the scatter ($\sigma$) is computed as the median absolute deviation divided by 0.6745, which is a robust measurement of scatter for small sample sizes \citep{Beers90}. 
\end{enumerate}

%an absolute calibration uncertainty of 2.5\% on average and better than 5.5\% in 95\% of the time.

To derive the final correction vector among the vectors of all the stars, we would like to take out the low-order difference among them but keep the high frequency variations in order to keep the constraining power on the high frequency mode. The high frequency variation is due to the telluric absorption by the atmosphere and should be the same among all stars. We would like to use the fact that different stars provides slightly different wavelength sampling. By combining them we can supersample the high frequency variation and provide a more accurate telluric correction.
To take out the lower-order difference, we first interpolate all vectors onto a common wavelength grid, and take the average among them. We then divide each vector by the average, and fit the result of this division by a 3rd-order polynomial function. These polynomial functions are a description of the low-order differences among all the correction vectors. We divide each original correction vector (in their original wavelength space) by their respective 3rd-order polynomial. The resulting vectors now have the same low-order shape but are still in their original wavelength grids, which are slightly different from one another. We call these the low-order-flattened calibration vectors. 

\begin{figure*}
\begin{center}
\includegraphics[width=1.0\textwidth, viewport =30 60 760 380, clip]{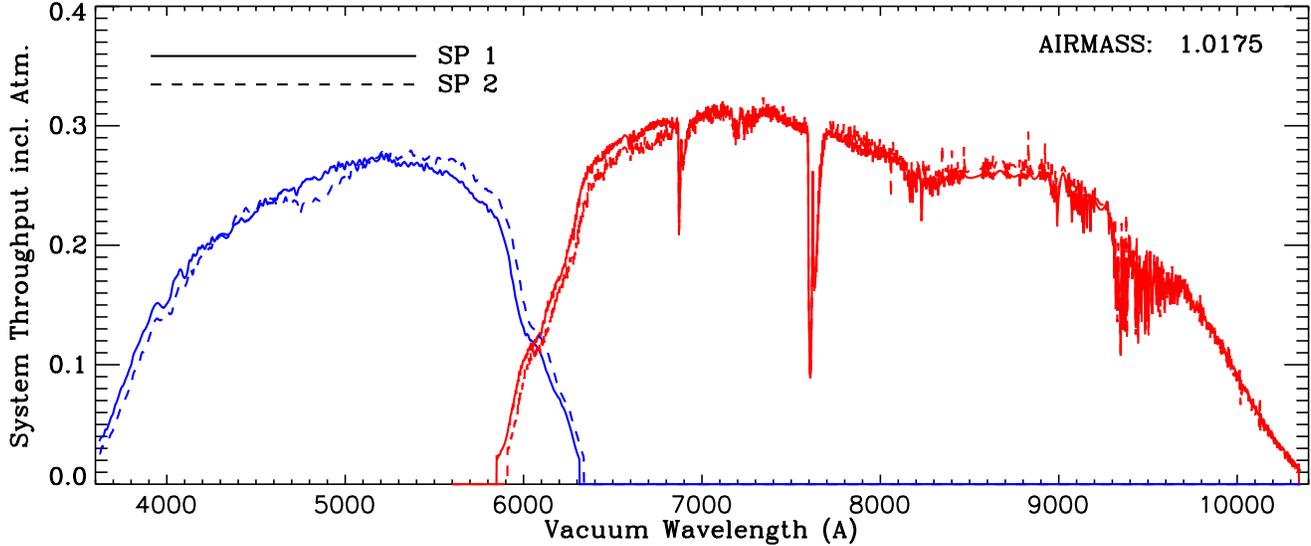}
\caption{The measured system throughput for one of the exposures (Exposure number: 177380) taken on MJD 56741 at an airmass of 1.0175. The throughput includes the transparency of the atmosphere, the efficiency of the whole telescope, an average fiber, spectrograph, and the detector. The solid and dashed curves show the throughput for Spectrograph 1 and 2, respectively, with the blue (left) and red (right) colors indicating the two cameras in each spectrograph.}
\label{fluxcal:fig:throughput}
\end{center}
\end{figure*}

The final step is to merge all of these low-order-flattened calibration vectors. With their slightly different wavelength grids, they supersample the spectral resolution element. We fit a b-spline to the merged spectrum with break points separated by $10n$-pixels in the blue (where $n$ is the number of good stars on a spectrograph) and  break points separated by $1.5n$-pixel in the red. The reason for the higher frequency fit in the red is to be able to sample the telluric absorption lines. Note the pixels here are much smaller than the original pixels because the merging of multiple spectra. The resulting `average' calibration vector is then applied to all of the other spectra in the same spectrograph and from the same exposure. Multiplying this calibration vector with the initial estimate of the throughput yields the final throughput curve of the system including atmospheric transparency. Examples of the derived throughput curves are shown in Figure~\ref{fluxcal:fig:throughput}. 

These curves can be compared with the throughput curves shown in Figure 38 of \cite{Smee13}, which are defined in the same way. Both the throughput shown here and those of \cite{Smee13} have made aperture corrections, but in different ways. \cite{Smee13} made the correction assuming a double Gaussian seeing profile with 1\arcsec\ FWHM (same for all wavelengths). The observations on which the BOSS throughput was based were conducted under seeing better than 1.15\arcsec\ and the four standard stars yielding the highest throughput were selected. Our mini-bundles provide much better aperture correction allowing us to derive accurate throughput from observations with much worse seeing. Our throughput is higher than BOSS's by a few percent in the blue and about 5\% in the red. This improvement is consistent with the expectation from our anti-reflection coatings \citep{Drory15}.

\section{Evaluating the Correction Accuracy}

\subsection{Consistency among independent measurements} \label{sec:consistency}

To evaluate the true calibration error as a function of wavelength, we check the consistency in the derived throughput vectors for different exposures, taken at different dither positions, and measured in different spectrographs. This guarantees the throughput vectors we are comparing are completely independent. Different exposures provide different PSF profiles, different dithers provide different sampling of the PSF, and different spectrographs provide different sets of standard stars. There is an intrinsic throughput difference between the two spectrographs but that is fixed for the fixed sets of fiber assemblies (different cartridges have small and negligible differences here). However, because our throughput vector includes the transparency of the atmosphere which varies constantly, the throughput comparison between any two exposures can include an intrinisic difference. To avoid this complication, we only look at pairs of consecutive exposures for which the transparency at the guider wavelength (measured from guider camera images in a broadband filter) differed by less than 3\%. This is satisfied by about 80\% of the exposure pairs. 

For each such exposure pair, we take the ratio between the throughput vector derived for Spectrograph 1 (SP1) from Exposure 1 and that for Spectrograph 2 (SP2) from Exposure 2. If each individual throughput vector has a fractional error of $x$, the ratio between the two would have a fractional error of $\sqrt{2} x$. 

\begin{figure*}
\begin{center}
\includegraphics[width=0.95\textwidth]{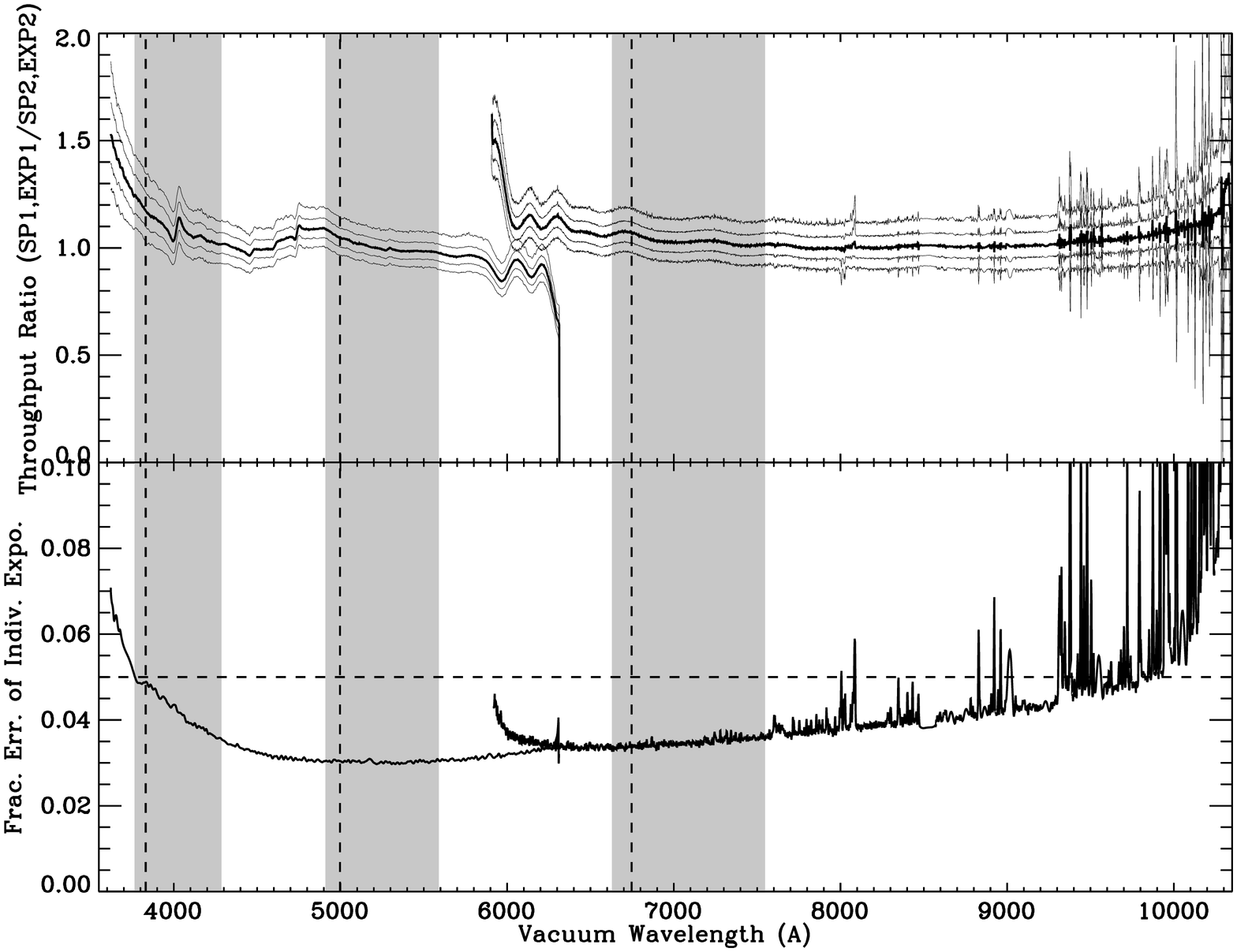}
\caption{Top: the throughput ratio distribution as a function of wavelength between two completely independent measurements of the throughput curves, constructed by dividing throughput measured in Spectrograph 1 for Exposure 1 and Spectrograph 2 for Exposure 2 for 627 consecutive exposure pairs with different dither positions but similar transparency. The thick line indicates the median ratio among these exposure pairs, which reflects the intrinsic throughput difference between the two spectrographs. The thin lines indicate the 2.5-, 15.85-, 84.15-, and 97.5-percentiles of the distribution at each wavelength, corresponding to the enclosed fractions of 1$\sigma$ and 2$\sigma$ limits of a Gaussian distribution. The curves below (above) 6000 \AA\ are for the blue (red) camera. Bottom: the estimated fractional error of the flux calibration for an individual exposure in a spectrograph. This is derived by taking the standard deviation of the 627 throughput ratio curves, divided by the mean ratio, then divided by $\sqrt{2}$. The horizontal line indicates the 5\% science requirement. For the great majority of the wavelengths, we achieved better than 5\% calibration for the random component (as opposed to the systematic component) of the absolute calibration. The grey bands indicate the positions of \oiiw, \hb\, and \hal\ (\niibw\ is close to \hal) for the redshift range of the MaNGA sample ($0.01<z<0.15$). The vertical dashed lines indicate their positions for the median redshift of MaNGA's Primary+ sample.
%Note this statistics is based on data from 6 different cartridges with different fiber assemblies. The different fibers would add to the variation.
}
\label{fig:spratios}
\end{center}
\end{figure*}

Figure~\ref{fig:spratios} shows the ratio vectors between the throughput vector pairs for 627 such exposure/spectrograph pairs. We always divide SP1 by SP2. The dark lines shows the median ratio at each wavelength, which reflects the intrinsic difference between the two spectrographs. The thinner dark lines show the 2.5-, 15.85-, 84.15-, and 97.5-percentiles of the distribution at each wavelength, corresponding to the enclosed fractions of 1$\sigma$ and 2$\sigma$ limits of a Gaussian distribution. The bottom panel shows the RMS of the fractional error divided by $\sqrt{2}$ to show the actual fractional error on each individual calibration. We achieve better than 5\% calibration for 89\% of the wavelength range. This is the random error component of the absolute calibration. 

This method also allows us to evaluate the relative calibration accuracy. For each throughput curve, we take the medians in two 20${\rm \AA}$-wide windows around \hb\ and \hal\ (redshifted to MaNGA sample's median redshift). The ratio between the two medians measures the relative calibration ($c_1$ in \S\ref{sec:requirements}) given by each throughput vector. We then divide the ratio from Spectrograph 1\ in Exposure 1 by the ratio from Spectrograph 2\ in Exposure 2. The resulting ratio has a fractional RMS scatter of 2.4\% among the 627 exposure pairs, corresponding to a 1.7\% fractional error on the relative calibration between \hal\ and \hb\ for each individual calibration vector. Doing the same calculation for \nii\ and \oii\ yields a 4.7\% fractional error on their relative calibration ($c_2$) for each individual calibration vector. These meet the science requirements specified in \S\ref{sec:requirements}. Given that the distribution of the spectrographs' throughput ratio is fairly close to a Gaussian distribution, these numbers correspond to roughly 68.3-percentile of the error distribution. 

%To make such evaluation, we again take the ratio of throughput curves between the two spectrographs for each exposure. Because the atmosphere condition could vary from exposure to exposure, directly comparing one exposure with another would involve this uncertainty. The two spectrographs always see the same condition, so their ratios should be fixed. Any deviation from the fixed ratio should be due to flux calibration error. For each plate, we group all exposures into three groups for the three dither positions. We take the median throughput ratio curve for each dither. Then at each wavelength, we take the min-to-max difference among the three dithers and divide that by the mean ratio among the three dithers. Thus for each plate, we arrive at a fractional min-to-max difference curve as a function of wavelength. For the 77 plates we examined, we show the scatter in Figure~\ref{fig:} which illustrate the 

\subsection{Comparison with broadband photometry}
\begin{figure*}
\begin{center}
\includegraphics[width=1.0\textwidth]{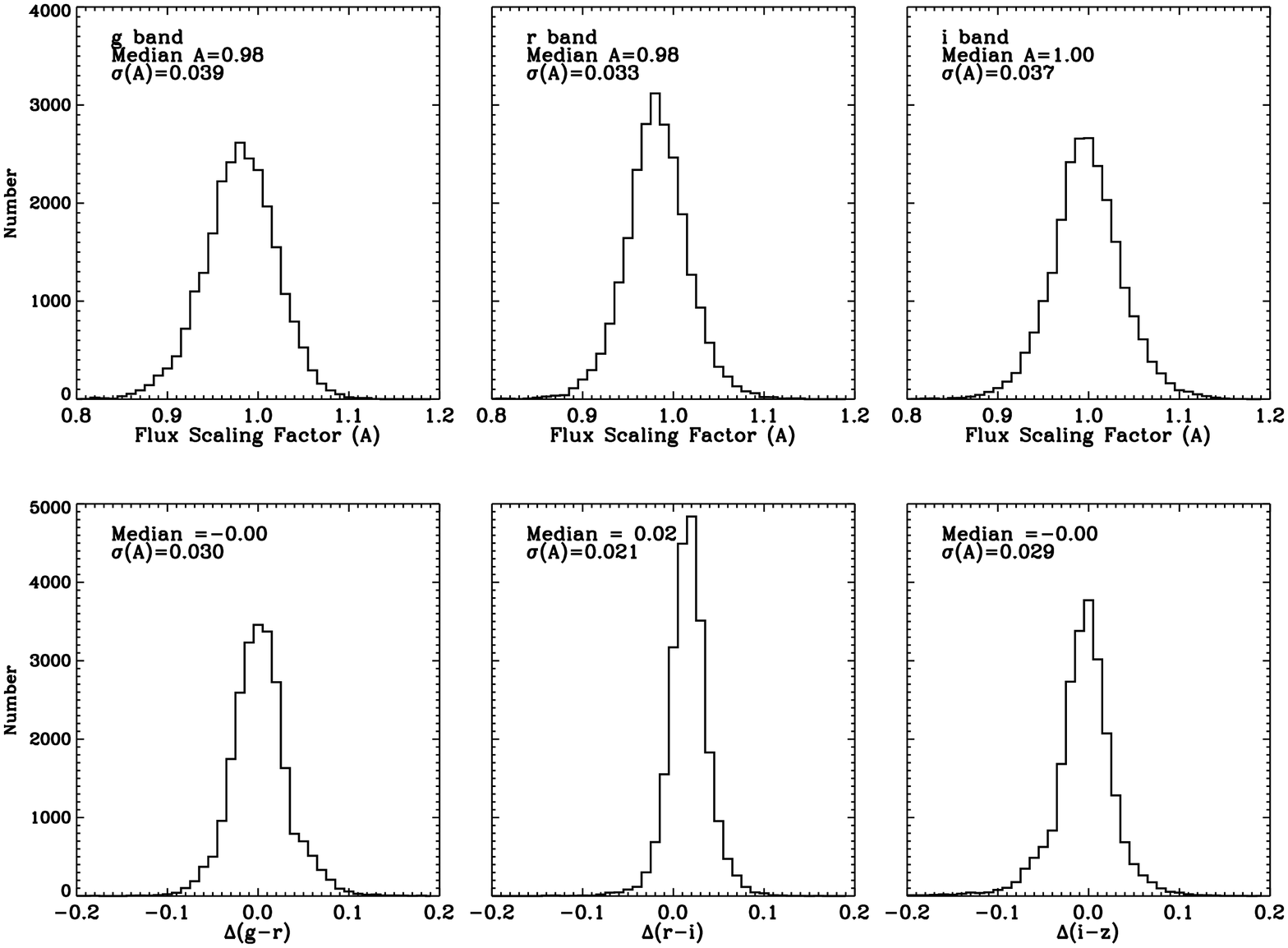}
\caption{Top: Distribution of flux scaling factor (A) derived through astrometry-matching MaNGA spectra to SDSS broadband images in $g$, $r$, and $i$ bands.
Bottom: Distribution of color residuals implied by the relative flux scaling ratios between different bands. }
%Each line represent one galaxy observed in one exposure. In total, 17 galaxies observed in 6 exposures are included. Some of the fitting results in large $\chi^2$ which also lead to large flux scaling error --- these are not reflecting error in flux calibration. The thick lines represent the median and $\pm1\sigma$ deviation among all fits with $\chi^2 < 1$. The dotted lines correspond to 2.5\% and 5\% deviations. }
\label{fig:fluxcal_broadband}
\end{center}
\end{figure*}

The above comparison provides a measurement of the random component of the calibration error, but it does not determine if there are any systematic offset across all exposures. In this section, we check our absolute accuracy of our spectrophotometric calibration by a comparison to SDSS photometry of galaxies. 
%Another indication of the calibration accuracy can be provided by a comparison to SDSS photometry. 
This comparison is done as part of the MaNGA Data Reduction Pipeline (Law et al. in prep). At a later stage in the pipeline, for each exposure, we register all the spectra taken for each galaxy to the image of that galaxy. Due to the finite mechanical tolerance between the fiber bundles and the holes on the plug plates, and due to imperfect guiding, there is uncertainty in the exact position and rotation of the fiber bundle relative to the galaxy for each exposure. Before we construct the data cube, we need to register the fiber spectra associated with each IFU in each exposure to the imaging. This is done in a manner similar to the method employed by the VENGA Survey \citep{Blanc13}. First, the synthetic broad-band flux of each fiber is computed by integrating the sky-subtracted, flux-calibrated spectra over the corresponding transmission curve. The code then explores a grid of offsets in position (RA, Dec) and rotation. At each position on this grid the fiber coordinates are shifted by the corresponding amounts and aperture photometry is performed on a PSF matched SDSS broad-band image using 2.0\arcsec\ diameter apertures at the corresponding position of each fiber. For example, for a 61-fiber bundle, there will be 61 synthetic $r$-band flux from MaNGA spectra and 61 $r$-band aperture photometry measurements from the image. The code then fits the MANGA synthetic flux of all fibers in a bundle against SDSS broad-band flux using the equation: $F_{\rm SDSS} = A\times F_{\rm MANGA} + B$. A perfect flux calibration and sky subtraction in both the spectra and the images would imply A=1 and B=0. This process yields an evaluation of the flux calibration accuracy for each galaxy in each exposure. Deviation of A from 1 indicate systematics in the absolute flux calibration relative to the imaging. Deviation of B from 0 indicate residuals in sky subtraction in either the imaging or the spectral data. Figure~\ref{fig:fluxcal_broadband} shows the distribution of the flux scaling factors (A) derived for all galaxies on the 64 plates observed before May 27th, 2015, with a total of 753 exposures, and 25,359 IFU-exposure combinations. Occasionally, this astrometry matching fails for reasons unrelated to flux calibration, which result in large $\chi ^2$. Here, we have removed the 5\% of cases where the $\chi ^2$ is larger than 3 indicating bad astrometry matching. The resulting absolute calibration accuracy is better than 4\% in all bands (upper panels in Figure~\ref{fig:fluxcal_broadband}) and the relative calibration between bands is better than 3\% (lower panels). This is well within the science requirements for MaNGA. We note that the median value for A is lower than 1 by 2\% for $g$-band and $r$-band, indicating a 2\% systematic difference between SDSS imaging calibration and our spectral data. The error could come from either the imaging or the spectral data, or both. Since we have met our science requirements, we do not try to sort out the source of the systematic difference here and leave it for future investigations.

%Note our flux calibration is derived and applied individually to each exposure. It shows the relative calibration across wavelength is good to a 2.5\% from 4500\AA\  to 8000\AA, with the maximum deviation happening towards the red (5\% at 9000\AA). This is well within the science requirements for MANGA of 2.5\% accuracy between \hal\ and \hb\ and 6\% accuracy between [NII] and [OII].  The few catstrophic failures all have large $\chi^2$ in the astrometry-matching and should not reflect flux calibration error. They mostly result from one IFU.

\section{Summary}
The science goals of integral field spectroscopy require that we strictly calibrate the spectrum of each aperture element in the IFU to the aperture flux density of a PSF-convolved surface brightness profile without artificially accounting for any missing flux due to aperture misalignment or atmospheric refraction. This is a more challenging goal than what is usually required for single-fiber spectroscopy science. 
%Integral field spectroscopy requires different spectrophotometry calibration technique from single-fiber spectroscopy. 
We have demonstrated the use of mini-bundles to achieve the separation of the flux loss due to throughput and flux loss due to a finite fiber aperture. The resulting {\it relative} calibration uncertainty has an RMS fractional error of 1.7\% between \hal\ and \hb, and 4.7\% between \nii\ and \oii. The absolute calibration is better than 5\% for 89\% of the wavelength range. These meet the science requirements for MaNGA. 
%meets the science requirement for MaNGA, with relative calibration between 6563\AA\ and 4861\AA\ to better than 2.5\%,. 

There are potential improvements we can make to further increase the accuracy of the spectrophotometry. These could include more detailed modeling of the asymmetric PSF, computing the actual guiding corrections to construct the time-integrated PSF, optimization to the PSF fitting procedure, improvements on the model grid and typing of the star. We could also use a large number of exposures to separate the time-invariant component of the correction vector from the time-dependent component. Since MaNGA science requirements are already met, we leave these ideas for future endeavors. 

%\section{TO DO}
%One essential thing is to look at the airmass dependence in the high-airmass plate and see if we need to use the airmass as the 2nd dimension in the bspline fitting. 

\acknowledgements

AW acknowledges support of a Leverhulme Trust Early Career Fellowship. DB acknowledges support by grant RSF 14-50-00043. 

This project made use of data taken in both SDSS-III and SDSS-IV. Funding for SDSS-III has been provided by the Alfred P.~Sloan Foundation, the Participating Institutions, the National Science Foundation, and the U.S.~Department of Energy Office of Science. Funding for the Sloan Digital Sky Survey IV has been provided by the Alfred P. Sloan Foundation, the U.S. Department of Energy Office of Science, and the Participating Institutions. SDSS-IV acknowledges support and resources from the Center for High-Performance Computing at the University of Utah. The SDSS web site is {\tt www.sdss.org} .

SDSS is managed by the Astrophysical Research Consortium for the Participating Institutions in both collaborations.  In SDSS-III
these include the University of Arizona, the Brazilian Participation Group, Brookhaven National Laboratory, Carnegie Mellon
University, University of Florida, the French Participation Group, the German Participation Group, Harvard University, the
Instituto de Astrofisica de Canarias, the Michigan State/Notre Dame/JINA Participation Group, Johns Hopkins University, Lawrence
Berkeley National Laboratory, Max Planck Institute for Astrophysics, Max Planck Institute for Extraterrestrial Physics, New Mexico
State University, New York University, Ohio State University, Pennsylvania State University, University of Portsmouth, Princeton
University, the Spanish Participation Group, University of Tokyo, University of Utah, Vanderbilt University, University of
Virginia, University of Washington, and Yale University.

The Participating Institutions in SDSS-IV include the Brazilian Participation Group, the Carnegie Institution for Science, Carnegie Mellon University, the Chilean Participation Group, the French Participation Group, Harvard-Smithsonian Center for Astrophysics, Instituto de Astrofísica de Canarias, The Johns Hopkins University, Kavli Institute for the Physics and Mathematics of the Universe (IPMU) / University of Tokyo, Lawrence Berkeley National Laboratory, Leibniz Institut für Astrophysik Potsdam (AIP), Max-Planck-Institut für Astronomie (MPIA Heidelberg), Max-Planck-Institut für Astrophysik (MPA Garching), Max-Planck-Institut für Extraterrestrische Physik (MPE), National Astronomical Observatory of China, New Mexico State University, New York University, University of Notre Dame, Observatório Nacional / MCTI, The Ohio State University, Pennsylvania State University, Shanghai Astronomical Observatory, United Kingdom Participation Group, Universidad Nacional Autónoma de México, University of Arizona, University of Colorado Boulder, University of Oxford, University of Portsmouth, University of Utah, University of Virginia, University of Washington, University of Wisconsin, Vanderbilt University, and Yale University.

\appendix
\section{A. Alternative calibration options considered}\label{sec:hardwarechoices}
%Besides the flux calibration technique adopted by MaNGA, we have also considered other options.
For MaNGA, we developed and tested several flux calibration methods. In this Appendix, we discuss other flux calibration options we considered but did not adopt, as these can potential be useful in other situations. 

As discussed in \S\ref{sec:principle}, we need to separate the throughput loss factor from any aperture-induced flux error. One concept was to use large fibers to get all the light from the star, which would be insensitive to differential atmosphere refraction at modest airmass and would suffer little aperture loss.
However, there are practical limitations on the fiber size due to the increasing stiffness of large fibers. Experiments
at Washburn Laboratories suggested that 5\arcsec\ (300 microns) was the largest fiber size that would be workable. With
moderately good seeing conditions of 1.3\arcsec\ and an airmass of 1.12, with 0.15\arcsec\ miscentering errors, a
5\arcsec\ fiber loses $<1\%$ of the PSF flux.

However, given that MaNGA's dither pattern traces an equilateral triangle 1.44\arcsec\ on a side, if we place the
center of a 5\arcsec\ fiber at the center of the dithering triangle, the light losses at different dither positions can
increase to as much as 10\% in the blue. To avoid this 10\% light loss, one has to use three sets of standard stars,
with each set designed for a different dither position. Considering the larger footprint of the fibers on the CCD, this would require a much larger allocation of our CCD real estate to calibration sources than our chosen method.

Another problem with this large-fiber scheme is that all of the IFU science fibers have a 2\arcsec\ core. The 5\arcsec\
fibers would therefore have a different spectral resolution. As a result, they would not be suitable for correcting 
high-frequency wavelength variations such as telluric absorption features. To achieve the telluric correction, one
would have to use both 5\arcsec\ and 2\arcsec\ fibers to target standard stars. Again, the total number of fibers
allocated for calibration becomes prohibitive.

\begin{figure}
\begin{center}
\includegraphics[width=1.0\textwidth,viewport=20 180 600 620,clip]{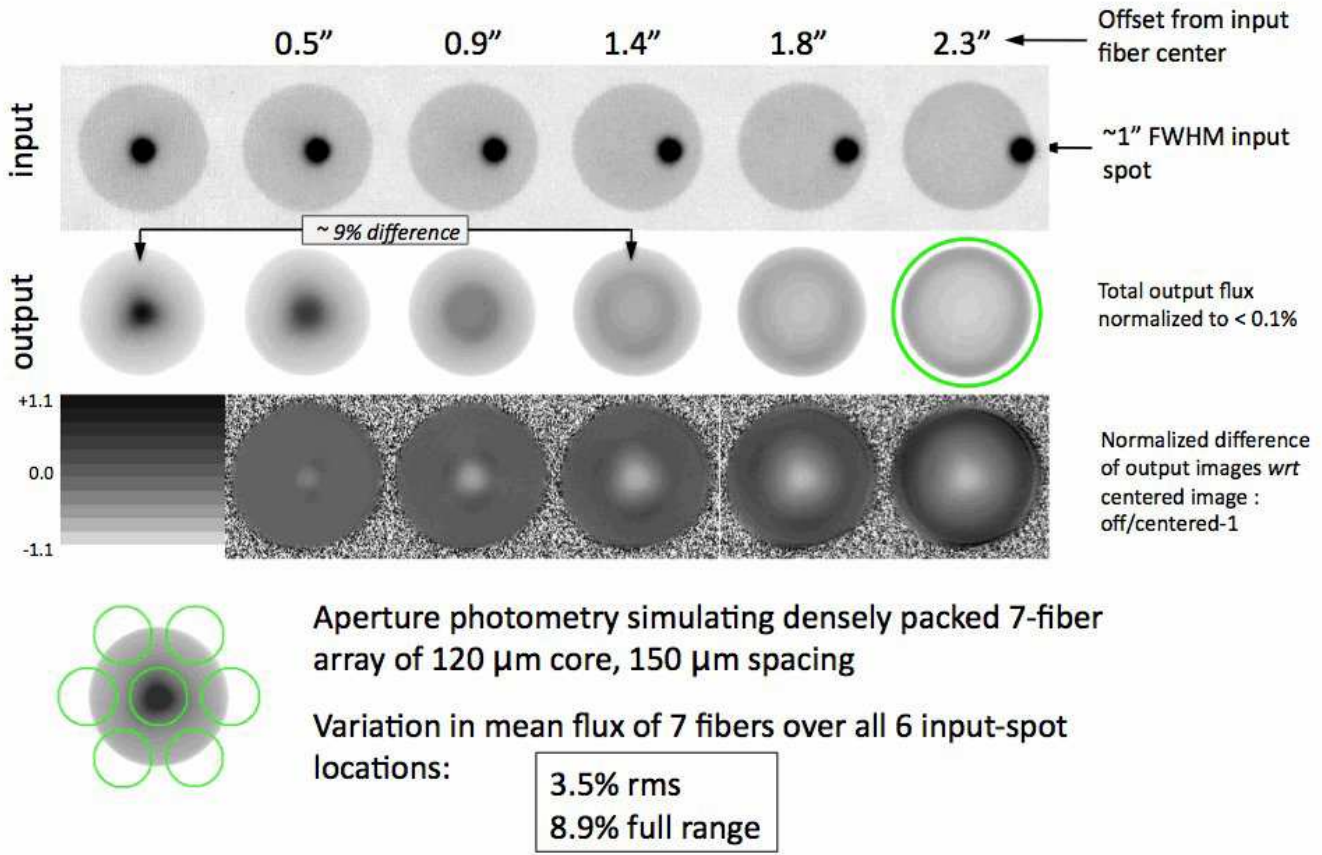}
\caption{Result of light scrambling inside a 2m-long fiber with a $300 \mu m$ core. The top row shows the six different
input beam location relative to the fiber. The middle row shows the output beams. The bottom row shows the normalized
difference of output images relative to the centered image. It shows the scrambling of light inside the 2m-long fiber is insufficient to always feed a smaller fiber with the same fraction of light, as the position of the star changes within the bundle due to dithered observations or DAR. 
Even if we couple an 7-fiber mini-bundle at the output end, the variation would still be at the several percent level. Additionally, DAR will make the fraction of light recovered by the 2\arcsec\ fibers a function of wavelength.}
\label{fluxcal:fig:scramble}
\end{center}
\end{figure}

\begin{figure}
\begin{center}
\includegraphics[width=0.7\textwidth,viewport=20 265 600 550,clip]{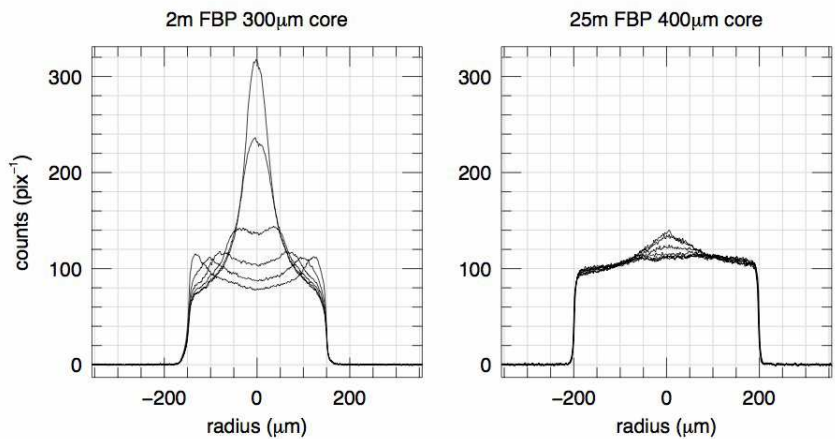}
\caption{The 1-d image profiles of the output beams as presented in Fig.~\ref{fluxcal:fig:scramble}. The two panels show the results of a 2m-long fiber with a $300\mu m$ core (left) and a 25m-long fiber with a $400\mu m$ core (right). The shorter fiber provides insufficient scrambling.}
\label{fluxcal:fig:crosscut}
\end{center}
\end{figure}

To measure the telluric correction at the same spectral resolution, another method we considered was to target standards with 5\arcsec\ fibers coupled to a single 2\arcsec\ fiber or a 7-fiber hexagonal bundles of 2\arcsec\ fibers, then feed the 2\arcsec\ fibers to the spectrographs. 
If the standard star light collected by the 5\arcsec\ fiber always emerged with a uniformly-illuminated beam, coupling the output beam to a single fiber or a 7-fiber bundle would yield the necessary flux information without a loss of resolution.
However, because the star will not illuminate the 5\arcsec\ fiber uniformly, given the short fiber length, the light will not be completely homogenized inside the 5\arcsec\ fiber. This is illustrated in Fig.~\ref{fluxcal:fig:scramble} and Fig.~\ref{fluxcal:fig:crosscut}. The scrambling of light inside a 2m-long fiber is insufficient to reduce monochromatic flux calibration uncertainties below 3.5\% level even if the output end were measured with a mini-bundle of 7 2\arcsec-fibers. Given the DAR, the resulting calibration would also have a wavelength-dependence at this level. Increasing the fiber length to 25m long would reduce the uncertainty to 0.3\%. However, the fiber throughput would be reduced significantly in the blue if the fibers are much longer than 2m (roughly a 20\% decrease in throughput at 400nm going from 2m to 25m fiber length). Therefore, MaNGA did not pursue this approach. The solution MaNGA adopted is presented in \S\ref{sec:implementation}

%The final method we investigated is to use the 7-fiber hexagonal bundle described above directly to target standard stars. Given the gaps between the fibers, not all the light will be collected. However, the relative flux ratios between the 7 fibers allow an accurate determination of the actual position of the star image inside the bundle. Given a priori knowledge about the PSF shape of the star obtained from the guider images and theoretical knowledge about differential atmosphere refraction, we can accurately reconstruct the fraction of light falling into each fiber at each wavelength. This allows us to estimate the aperture loss separately from the throughput loss. %A similar technique was applied in the VENGA survey \citep{Blanc09} 

\section{B. Comparison with a traditional spectrophotometric standard star}
In this appendix, we illustrate our calibration accuracy using a traditional spectrophotometirc standard star observed with our system. One should keep in mind that this is a single data point so it cannot be used to establish the statistics of our calibration accuracy. Nonetheless, it is useful to check. 
%The above comparison provides a measurement of the random component of the calibration error, but it does not determine if there are any systematic offset across all exposures. In this section, we check our absolute spectrophotometric calibration using an external calibrator. 
During our commissioning observation, we placed a bundle on the standard star HZ 21 for this purpose. 

We would also like to note that producing a spectrum for a star using our setup requires an additional step than producing a calibrated spectrum for a fiber in a bundle. For our galaxy targets, we just need to apply the average throughput correction. For a star, we also need to know the PSF-covering fraction for that specific star. Because the individual PSF-covering fraction as a function of wavelength is more uncertain than the average of many stars, the result could only be worse than the actual flux calibration accuracy we achieve in the galaxy data. Nonetheless, this provides a conservative indication of our absolute calibration error. 

%It is important to note that this is {\it not} the best way to evaluate the absolute performance of the flux calibration. Because what we derived is the correction vector for surface spectrophotometry, it is not easily applicable to a point source. In deriving the spectrum for a standard star, we not only have to derive the correction vector that would be applied to all science spectra, we also need to know the PSF-covering fraction derived for the specific star. 

\begin{figure*}
\begin{center}
\includegraphics[width=1.0\textwidth, viewport= 0 30 760 620,clip]{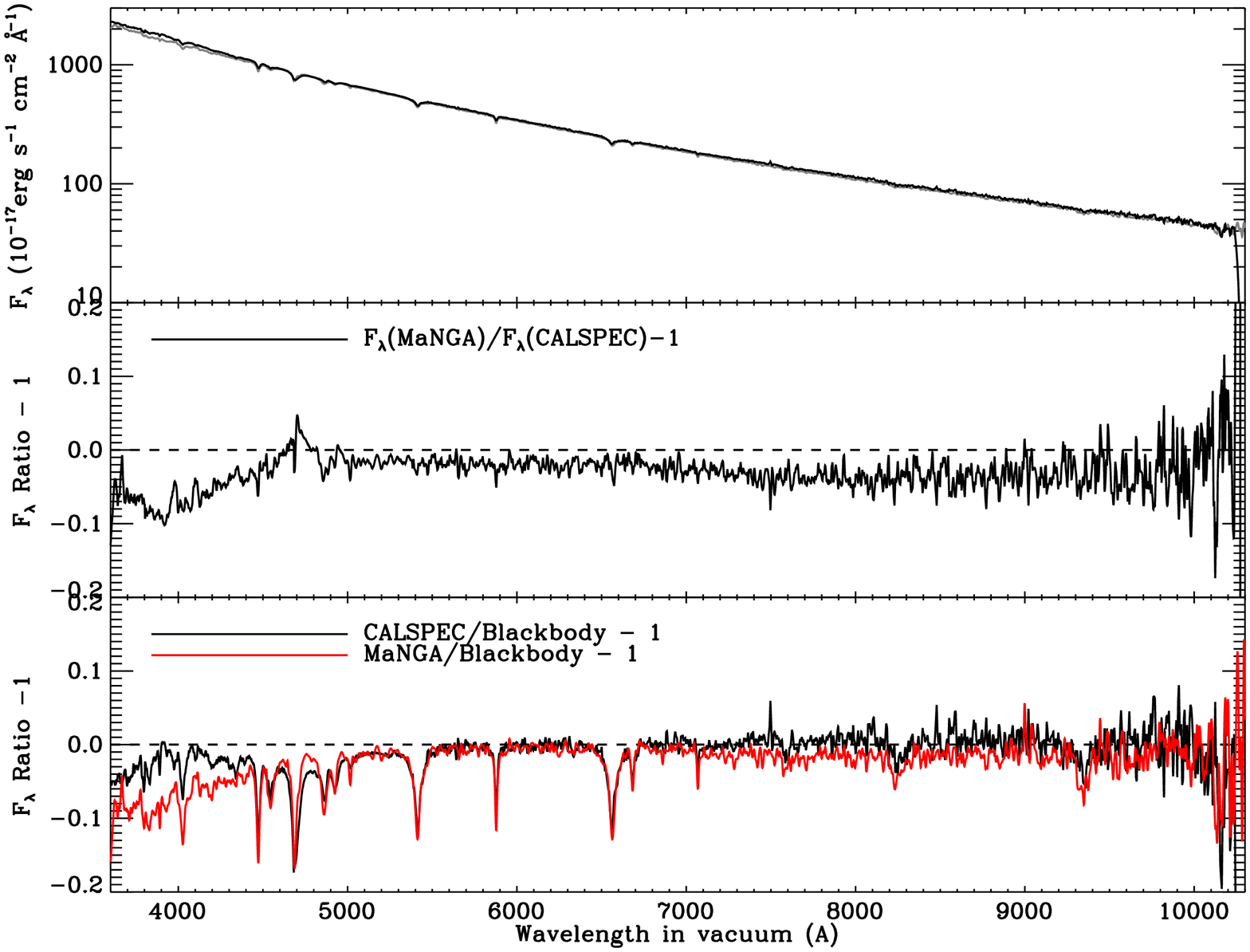}
\caption{Upper panel: Spectra of an Oke standard star, HZ 21, as given by the CALSPEC database (dark black line) and that obtained from the average of 6 MaNGA exposures (gray line). The systematic error is very small, which is detailed in the bottom panel. Middle panel: Fractional deviation of the average derived spectrum for HZ 21 from 6 exposures, after convolving to the resolution of the CALSPEC spectrum. 
%Note, this comparison includes the error on the PSF-covering of an individual star in addition to the error on the absolute calibration for galaxies. Thus, it could be worse than the actual calibration error of galaxy spectra. 
Bottom panel: Fractional deviations of our HZ 21 spectrum and CALSPEC spectrum relative to a blackbody spectrum with T=100,000K normalized around 6000-6100\AA. This shows both our spectrum and CALSPEC spectrum have residual systematic errors in the blue wavelengths at a level of 5-10\%. Our spectrum is better than CALSPEC between 4600-5000\AA\ but worse below 4600\AA. Note this observation of one star does not fully reflect the statistical accuracy of our flux calibration.
}
\label{fig:hz21}
\end{center}
\end{figure*}

We compare our derived HZ 21 spectra with the standard spectrum given by STScI's CALSPEC database\footnote{http://www.stsci.edu/hst/observatory/crds/calspec.html}, which is derived by combining HST/STIS observations with spectra taken by J. Oke \citep{Oke90, Bohlin01, Bohlin07}. Figure~\ref{fig:hz21} (top panel) shows the standard spectrum given by the CALSPEC database and the average spectrum we obtained from six dithered exposures. The two spectra trace each other fairly well. It is worth noting our groundbased spectrum have very good telluric correction that it is as smooth in telluric regions as the HST/STIS spectrum which is not affected by telluric features. Our spectrum has a much higher spectral resolution as one could see from the depth of many lines, such as \heiilw. In the middle panel, we convolved our spectra to the resolution as given by CALSPEC, then derived the fractional deviation from the CALSPEC spectrum. The residual show some large scale tilt in certain wavelength windows and some small-scale features. It is quite plausible that there are systematics in the CALSPEC spectrm at this level as well. Since a hot white dwarf's spectrum is very close to a blackbody,  we can compare both spectra to a blackbody spectrum to check the systematic error in them. In the literature, there are discrepant measurements for the effective temperature for HZ 21, ranging from around 50,000K \citep{Koester79} to 100,000K \citep{OkeS71, Reynolds03}. Both our spectrum and CALSPEC spectrum agree much better with a 100,000K blackbody spectrum for wavelengths redder than 5000\AA. In the bottom panel of Figure\ref{fig:hz21}, we divide both spectra by a 100,000K blackbody that has been normalized to each spectrum between 6000\AA-6100\AA. Since the blackbody spectrum has no absorption lines, only the line-free regions reflect the residual systematics in the data.

This comparison indicates both our spectrum and the CALSPEC spectrum have some small systematics. CALSPEC spectrum has a very broad dip between 4000 and 5000\AA\ with a 5\% maximum deviation, and a dip below 3900\AA. The part of the CALSPEC spectrum blueward of 4683\AA\ is from \cite{Oke90} and is stitched together with HST/STIS spectrum at 4683\AA\ around the line center of \heiilw. Our spectrum shows a tilt blueward of 4600\AA\ that goes down to -10\% at 3800\AA\ and a slight tilt redward of 6800\AA of 1-2\%. 
We suspect these systematics could originate from the error in the derived PSF-covering fractions for the F-star standards on this plate and that for HZ 21, which can be due to the simplified assumptions we make about the PSF regarding its circular symmetry and how it changes with focus offset. Given our reported statistics in \S\ref{sec:consistency}, the systematics shown in this single spectrum is generally within $1\sigma$, and at $\sim2\sigma$ at the worst part.

%The feature at 4700\AA\ is very likely due to a problem in the CALSPEC spectrum. Overall, the residual deviation is fairly small at most wavelengths and has very little wavelength dependence, except below 5000\AA\ where we see a wavelength dependence and the residual could reach 10\%. We suspect some of this error could be due to the error in the derived PSF-covering fraction of this individual star, rather than the calibration systematics of our method. The evidence for this conclusion comes from comparisons to broadband galaxy photometry, which we discuss below. It is worth noting this comparison is a single data point and cannot represent the average calibration accuracy we achieved.

\bibliographystyle{apj}
\bibliography{mangaspecphot_references}

\begin{thebibliography}{33}
\expandafter\ifx\csname natexlab\endcsname\relax\def\natexlab#1{#1}\fi

\bibitem[{{Adelman-McCarthy} {et~al.}(2008)}]{SDSSDR6}
{Adelman-McCarthy}, J.~K., {et~al.} 2008, \apjs, 175, 297

\bibitem[{{Allen} {et~al.}(2015){Allen}, {Croom}, {Konstantopoulos}, {Bryant},
  {Sharp}, {Cecil}, {Fogarty}, {Foster}, {Green}, {Ho}, {Owers}, {Schaefer},
  {Scott}, {Bauer}, {Baldry}, {Barnes}, {Bland-Hawthorn}, {Bloom}, {Brough},
  {Colless}, {Cortese}, {Couch}, {Drinkwater}, {Driver}, {Goodwin},
  {Gunawardhana}, {Hampton}, {Hopkins}, {Kewley}, {Lawrence}, {Leon-Saval},
  {Liske}, {L{\'o}pez-S{\'a}nchez}, {Lorente}, {McElroy}, {Medling}, {Mould},
  {Norberg}, {Parker}, {Power}, {Pracy}, {Richards}, {Robotham}, {Sweet},
  {Taylor}, {Thomas}, {Tonini}, \& {Walcher}}]{Allen15}
{Allen}, J.~T., {et~al.} 2015, \mnras, 446, 1567

\bibitem[{{Beers} {et~al.}(1990){Beers}, {Flynn}, \& {Gebhardt}}]{Beers90}
{Beers}, T.~C., {Flynn}, K., \& {Gebhardt}, K. 1990, \aj, 100, 32

\bibitem[{{Bershady} {et~al.}(2010){Bershady}, {Verheijen}, {Swaters},
  {Andersen}, {Westfall}, \& {Martinsson}}]{Bershady10}
{Bershady}, M.~A., {Verheijen}, M.~A.~W., {Swaters}, R.~A., {Andersen}, D.~R.,
  {Westfall}, K.~B., \& {Martinsson}, T. 2010, \apj, 716, 198

\bibitem[{{Blanc} {et~al.}(2013){Blanc}, {Weinzirl}, {Song}, {Heiderman},
  {Gebhardt}, {Jogee}, {Evans}, {van den Bosch}, {Luo}, {Drory}, {Fabricius},
  {Fisher}, {Hao}, {Kaplan}, {Marinova}, {Vutisalchavakul}, \&
  {Yoachim}}]{Blanc13}
{Blanc}, G.~A., {et~al.} 2013, \aj, 145, 138

\bibitem[{{Bohlin}(2007)}]{Bohlin07}
{Bohlin}, R.~C. 2007, in Astronomical Society of the Pacific Conference Series,
  Vol. 364, The Future of Photometric, Spectrophotometric and Polarimetric
  Standardization, ed. C.~{Sterken}, 315

\bibitem[{{Bohlin} {et~al.}(2001){Bohlin}, {Dickinson}, \&
  {Calzetti}}]{Bohlin01}
{Bohlin}, R.~C., {Dickinson}, M.~E., \& {Calzetti}, D. 2001, \aj, 122, 2118

\bibitem[{{Boyd}(1978)}]{Boyd78}
{Boyd}, R.~W. 1978, Journal of the Optical Society of America (1917-1983), 68,
  877

\bibitem[{{Bryant} {et~al.}(2014){Bryant}, {Owers}, {Robotham}, {Croom},
  {Driver}, {Drinkwater}, {Lorente}, {Cortese}, {Scott}, {Colless}, {Schaefer},
  {Taylor}, {Konstantopoulos}, {Allen}, {Baldry}, {Barnes}, {Bauer},
  {Bland-Hawthorn}, {Bloom}, {Brooks}, {Brough}, {Cecil}, {Couch}, {Croton},
  {Davies}, {Ellis}, {Fogarty}, {Foster}, {Glazebrook}, {Goodwin}, {Green},
  {Gunawardhana}, {Hampton}, {Ho}, {Hopkins}, {Kewley}, {Lawrence},
  {Leon-Saval}, {Leslie}, {Lewis}, {Liske}, {Lopez-Sanchez}, {Mahajan},
  {Medling}, {Metcalfe}, {Meyer}, {Mould}, {Obreschkow}, {O'Toole}, {Pracy},
  {Richards}, {Shanks}, {Sharp}, {Sweet}, {Thomas}, {Tonini}, \&
  {Walcher}}]{Bryant14}
{Bryant}, J.~J., {et~al.} 2014, ArXiv e-prints

\bibitem[{{Bundy} {et~al.}(2015){Bundy}, {Bershady}, {Law}, {Yan}, {Drory},
  {MacDonald}, {Wake}, {Cherinka}, {S{\'a}nchez-Gallego}, {Weijmans}, {Thomas},
  {Tremonti}, {Masters}, {Coccato}, {Diamond-Stanic}, {Arag{\'o}n-Salamanca},
  {Avila-Reese}, {Badenes}, {Falc{\'o}n-Barroso}, {Belfiore}, {Bizyaev},
  {Blanc}, {Bland-Hawthorn}, {Blanton}, {Brownstein}, {Byler}, {Cappellari},
  {Conroy}, {Dutton}, {Emsellem}, {Etherington}, {Frinchaboy}, {Fu}, {Gunn},
  {Harding}, {Johnston}, {Kauffmann}, {Kinemuchi}, {Klaene}, {Knapen},
  {Leauthaud}, {Li}, {Lin}, {Maiolino}, {Malanushenko}, {Malanushenko}, {Mao},
  {Maraston}, {McDermid}, {Merrifield}, {Nichol}, {Oravetz}, {Pan}, {Parejko},
  {Sanchez}, {Schlegel}, {Simmons}, {Steele}, {Steinmetz}, {Thanjavur},
  {Thompson}, {Tinker}, {van den Bosch}, {Westfall}, {Wilkinson}, {Wright},
  {Xiao}, \& {Zhang}}]{Bundy15}
{Bundy}, K., {et~al.} 2015, \apj, 798, 7

\bibitem[{{Cappellari} {et~al.}(2011){Cappellari}, {Emsellem}, {Krajnovi{\'c}},
  {McDermid}, {Scott}, {Verdoes Kleijn}, {Young}, {Alatalo}, {Bacon}, {Blitz},
  {Bois}, {Bournaud}, {Bureau}, {Davies}, {Davis}, {de Zeeuw}, {Duc},
  {Khochfar}, {Kuntschner}, {Lablanche}, {Morganti}, {Naab}, {Oosterloo},
  {Sarzi}, {Serra}, \& {Weijmans}}]{Cappellari11}
{Cappellari}, M., {et~al.} 2011, \mnras, 413, 813

\bibitem[{{Dawson} {et~al.}(2013){Dawson}, {Schlegel}, {Ahn}, {Anderson},
  {Aubourg}, {Bailey}, {Barkhouser}, {Bautista}, {Beifiori}, {Berlind},
  {Bhardwaj}, {Bizyaev}, {Blake}, {Blanton}, {Blomqvist}, {Bolton}, {Borde},
  {Bovy}, {Brandt}, {Brewington}, {Brinkmann}, {Brown}, {Brownstein}, {Bundy},
  {Busca}, {Carithers}, {Carnero}, {Carr}, {Chen}, {Comparat}, {Connolly},
  {Cope}, {Croft}, {Cuesta}, {da Costa}, {Davenport}, {Delubac}, {de Putter},
  {Dhital}, {Ealet}, {Ebelke}, {Eisenstein}, {Escoffier}, {Fan}, {Filiz Ak},
  {Finley}, {Font-Ribera}, {G{\'e}nova-Santos}, {Gunn}, {Guo}, {Haggard},
  {Hall}, {Hamilton}, {Harris}, {Harris}, {Ho}, {Hogg}, {Holder}, {Honscheid},
  {Huehnerhoff}, {Jordan}, {Jordan}, {Kauffmann}, {Kazin}, {Kirkby}, {Klaene},
  {Kneib}, {Le Goff}, {Lee}, {Long}, {Loomis}, {Lundgren}, {Lupton}, {Maia},
  {Makler}, {Malanushenko}, {Malanushenko}, {Mandelbaum}, {Manera}, {Maraston},
  {Margala}, {Masters}, {McBride}, {McDonald}, {McGreer}, {McMahon}, {Mena},
  {Miralda-Escud{\'e}}, {Montero-Dorta}, {Montesano}, {Muna}, {Myers},
  {Naugle}, {Nichol}, {Noterdaeme}, {Nuza}, {Olmstead}, {Oravetz}, {Oravetz},
  {Owen}, {Padmanabhan}, {Palanque-Delabrouille}, {Pan}, {Parejko},
  {P{\^a}ris}, {Percival}, {P{\'e}rez-Fournon}, {P{\'e}rez-R{\`a}fols},
  {Petitjean}, {Pfaffenberger}, {Pforr}, {Pieri}, {Prada}, {Price-Whelan},
  {Raddick}, {Rebolo}, {Rich}, {Richards}, {Rockosi}, {Roe}, {Ross}, {Ross},
  {Rossi}, {Rubi{\~n}o-Martin}, {Samushia}, {S{\'a}nchez}, {Sayres}, {Schmidt},
  {Schneider}, {Sc{\'o}ccola}, {Seo}, {Shelden}, {Sheldon}, {Shen}, {Shu},
  {Slosar}, {Smee}, {Snedden}, {Stauffer}, {Steele}, {Strauss}, {Streblyanska},
  {Suzuki}, {Swanson}, {Tal}, {Tanaka}, {Thomas}, {Tinker}, {Tojeiro},
  {Tremonti}, {Vargas Maga{\~n}a}, {Verde}, {Viel}, {Wake}, {Watson}, {Weaver},
  {Weinberg}, {Weiner}, {West}, {White}, {Wood-Vasey}, {Yeche}, {Zehavi},
  {Zhao}, \& {Zheng}}]{Dawson13}
{Dawson}, K.~S., {et~al.} 2013, \aj, 145, 10

\bibitem[{{de Zeeuw} {et~al.}(2002){de Zeeuw}, {Bureau}, {Emsellem}, {Bacon},
  {Carollo}, {Copin}, {Davies}, {Kuntschner}, {Miller}, {Monnet}, {Peletier},
  \& {Verolme}}]{deZeeuw02}
{de Zeeuw}, P.~T., {et~al.} 2002, \mnras, 329, 513

\bibitem[{{Drory} {et~al.}(2015){Drory}, {MacDonald}, {Bershady}, {Bundy},
  {Gunn}, {Law}, {Smith}, {Stoll}, {Tremonti}, {Wake}, {Yan}, {Weijmans},
  {Byler}, {Cherinka}, {Cope}, {Eigenbrot}, {Harding}, {Holder}, {Huehnerhoff},
  {Jaehnig}, {Jansen}, {Klaene}, {Paat}, {Percival}, \& {Sayres}}]{Drory15}
{Drory}, N., {et~al.} 2015, \aj, 149, 77

\bibitem[{{Fried}(1966)}]{Fried66}
{Fried}, D.~L. 1966, Journal of the Optical Society of America (1917-1983), 56,
  1372

\bibitem[{{Garc{\'{\i}}a-Benito} {et~al.}(2015){Garc{\'{\i}}a-Benito},
  {Zibetti}, {S{\'a}nchez}, {Husemann}, {de Amorim}, {Castillo-Morales}, {Cid
  Fernandes}, {Ellis}, {Falc{\'o}n-Barroso}, {Galbany}, {Gil de Paz},
  {Gonz{\'a}lez Delgado}, {Lacerda}, {L{\'o}pez-Fernandez}, {de
  Lorenzo-C{\'a}ceres}, {Lyubenova}, {Marino}, {Mast}, {Mendoza}, {P{\'e}rez},
  {Vale Asari}, {Aguerri}, {Ascasibar}, {Bekerait*error*{\.e}},
  {Bland-Hawthorn}, {Barrera-Ballesteros}, {Bomans}, {Cano-D{\'{\i}}az},
  {Catal{\'a}n-Torrecilla}, {Cortijo}, {Delgado-Inglada}, {Demleitner},
  {Dettmar}, {D{\'{\i}}az}, {Florido}, {Gallazzi}, {Garc{\'{\i}}a-Lorenzo},
  {Gomes}, {Holmes}, {Iglesias-P{\'a}ramo}, {Jahnke}, {Kalinova}, {Kehrig},
  {Kennicutt}, {L{\'o}pez-S{\'a}nchez}, {M{\'a}rquez}, {Masegosa}, {Meidt},
  {Mendez-Abreu}, {Moll{\'a}}, {Monreal-Ibero}, {Morisset}, {del Olmo},
  {Papaderos}, {P{\'e}rez}, {Quirrenbach}, {Rosales-Ortega}, {Roth},
  {Ruiz-Lara}, {S{\'a}nchez-Bl{\'a}zquez}, {S{\'a}nchez-Menguiano}, {Singh},
  {Spekkens}, {Stanishev}, {Torres-Papaqui}, {van de Ven}, {Vilchez},
  {Walcher}, {Wild}, {Wisotzki}, {Ziegler}, {Alves}, {Barrado}, {Quintana}, \&
  {Aceituno}}]{GarciaBenito15}
{Garc{\'{\i}}a-Benito}, R., {et~al.} 2015, \aap, 576, A135

\bibitem[{{Gunn} {et~al.}(2006){Gunn}, {Siegmund}, {Mannery}, {Owen}, {Hull},
  {Leger}, {Carey}, {Knapp}, {York}, {Boroski}, {Kent}, {Lupton}, {Rockosi},
  {Evans}, {Waddell}, {Anderson}, {Annis}, {Barentine}, {Bartoszek}, {Bastian},
  {Bracker}, {Brewington}, {Briegel}, {Brinkmann}, {Brown}, {Carr},
  {Czarapata}, {Drennan}, {Dombeck}, {Federwitz}, {Gillespie}, {Gonzales},
  {Hansen}, {Harvanek}, {Hayes}, {Jordan}, {Kinney}, {Klaene}, {Kleinman},
  {Kron}, {Kresinski}, {Lee}, {Limmongkol}, {Lindenmeyer}, {Long}, {Loomis},
  {McGehee}, {Mantsch}, {Neilsen}, {Neswold}, {Newman}, {Nitta}, {Peoples},
  {Pier}, {Prieto}, {Prosapio}, {Rivetta}, {Schneider}, {Snedden}, \&
  {Wang}}]{Gunn06}
{Gunn}, J.~E., {et~al.} 2006, \aj, 131, 2332

\bibitem[{{Kennicutt}(1998)}]{Kennicutt98}
{Kennicutt}, R.~C. 1998, \araa, 36, 189

\bibitem[{{Kewley} \& {Dopita}(2002)}]{KewleyD02}
{Kewley}, L.~J., \& {Dopita}, M.~A. 2002, \apjs, 142, 35

\bibitem[{{Koester} {et~al.}(1979){Koester}, {Liebert}, \& {Hege}}]{Koester79}
{Koester}, D., {Liebert}, J., \& {Hege}, E.~K. 1979, \aap, 71, 163

\bibitem[{{Law} {et~al.}(2015){Law}, {Yan}, {Bershady}, {Bundy}, {Cherinka},
  {Drory}, {MacDonald}, {S{\'a}nchez-Gallego}, {Wake}, {Weijmans}, {Blanton},
  {Klaene}, {Moran}, {Sanchez}, \& {Zhang}}]{Law15}
{Law}, D.~R., {et~al.} 2015, \aj, 150, 19

\bibitem[{{Margala} {et~al.}(2015){Margala}, {Kirkby}, {Dawson}, {Bailey},
  {Blanton}, \& {Schneider}}]{Margala15}
{Margala}, D., {Kirkby}, D., {Dawson}, K., {Bailey}, S., {Blanton}, M., \&
  {Schneider}, D.~P. 2015, ArXiv e-prints

\bibitem[{{O'Donnell}(1994)}]{O'Donnell94}
{O'Donnell}, J.~E. 1994, \apj, 422, 158

\bibitem[{{Oke}(1990)}]{Oke90}
{Oke}, J.~B. 1990, \aj, 99, 1621

\bibitem[{{Oke} \& {Shipman}(1971)}]{OkeS71}
{Oke}, J.~B., \& {Shipman}, H.~L. 1971, in IAU Symposium, Vol.~42, White
  Dwarfs, ed. W.~J. {Luyten}, 67

\bibitem[{{Osterbrock} \& {Ferland}(2006)}]{OsterbrockBook}
{Osterbrock}, D.~E., \& {Ferland}, G.~J. 2006, {Astrophysics of gaseous nebulae
  and active galactic nuclei}

\bibitem[{{Reynolds} {et~al.}(2003){Reynolds}, {de Bruijne}, {Perryman},
  {Peacock}, \& {Bridge}}]{Reynolds03}
{Reynolds}, A.~P., {de Bruijne}, J.~H.~J., {Perryman}, M.~A.~C., {Peacock}, A.,
  \& {Bridge}, C.~M. 2003, \aap, 400, 1209

\bibitem[{{Rosales-Ortega} {et~al.}(2010){Rosales-Ortega}, {Kennicutt},
  {S{\'a}nchez}, {D{\'{\i}}az}, {Pasquali}, {Johnson}, \&
  {Hao}}]{Rosales-Ortega10}
{Rosales-Ortega}, F.~F., {Kennicutt}, R.~C., {S{\'a}nchez}, S.~F.,
  {D{\'{\i}}az}, A.~I., {Pasquali}, A., {Johnson}, B.~D., \& {Hao}, C.~N. 2010,
  \mnras, 405, 735

\bibitem[{{S{\'a}nchez} {et~al.}(2012){S{\'a}nchez}, {Kennicutt}, {Gil de Paz},
  {van de Ven}, {V{\'{\i}}lchez}, {Wisotzki}, {Walcher}, {Mast}, {Aguerri},
  {Albiol-P{\'e}rez}, {Alonso-Herrero}, {Alves}, {Bakos}, {Bart{\'a}kov{\'a}},
  {Bland-Hawthorn}, {Boselli}, {Bomans}, {Castillo-Morales}, {Cortijo-Ferrero},
  {de Lorenzo-C{\'a}ceres}, {Del Olmo}, {Dettmar}, {D{\'{\i}}az}, {Ellis},
  {Falc{\'o}n-Barroso}, {Flores}, {Gallazzi}, {Garc{\'{\i}}a-Lorenzo},
  {Gonz{\'a}lez Delgado}, {Gruel}, {Haines}, {Hao}, {Husemann},
  {Igl{\'e}sias-P{\'a}ramo}, {Jahnke}, {Johnson}, {Jungwiert}, {Kalinova},
  {Kehrig}, {Kupko}, {L{\'o}pez-S{\'a}nchez}, {Lyubenova}, {Marino},
  {M{\'a}rmol-Queralt{\'o}}, {M{\'a}rquez}, {Masegosa}, {Meidt},
  {Mendez-Abreu}, {Monreal-Ibero}, {Montijo}, {Mour{\~a}o}, {Palacios-Navarro},
  {Papaderos}, {Pasquali}, {Peletier}, {P{\'e}rez}, {P{\'e}rez}, {Quirrenbach},
  {Rela{\~n}o}, {Rosales-Ortega}, {Roth}, {Ruiz-Lara},
  {S{\'a}nchez-Bl{\'a}zquez}, {Sengupta}, {Singh}, {Stanishev}, {Trager},
  {Vazdekis}, {Viironen}, {Wild}, {Zibetti}, \& {Ziegler}}]{Sanchez12}
{S{\'a}nchez}, S.~F., {et~al.} 2012, \aap, 538, A8

\bibitem[{{Schlegel} {et~al.}(1998){Schlegel}, {Finkbeiner}, \&
  {Davis}}]{SchlegelFD98}
{Schlegel}, D.~J., {Finkbeiner}, D.~P., \& {Davis}, M. 1998, \apj, 500, 525

\bibitem[{{Sharp} {et~al.}(2015){Sharp}, {Allen}, {Fogarty}, {Croom},
  {Cortese}, {Green}, {Nielsen}, {Richards}, {Scott}, {Taylor}, {Barnes},
  {Bauer}, {Birchall}, {Bland-Hawthorn}, {Bloom}, {Brough}, {Bryant}, {Cecil},
  {Colless}, {Couch}, {Drinkwater}, {Driver}, {Foster}, {Goodwin},
  {Gunawardhana}, {Ho}, {Hampton}, {Hopkins}, {Jones}, {Konstantopoulos},
  {Lawrence}, {Leslie}, {Lewis}, {Liske}, {L{\'o}pez-S{\'a}nchez}, {Lorente},
  {McElroy}, {Medling}, {Mahajan}, {Mould}, {Parker}, {Pracy}, {Obreschkow},
  {Owers}, {Schaefer}, {Sweet}, {Thomas}, {Tonini}, \& {Walcher}}]{Sharp15}
{Sharp}, R., {et~al.} 2015, \mnras, 446, 1551

\bibitem[{{Smee} {et~al.}(2013){Smee}, {Gunn}, {Uomoto}, {Roe}, {Schlegel},
  {Rockosi}, {Carr}, {Leger}, {Dawson}, {Olmstead}, {Brinkmann}, {Owen},
  {Barkhouser}, {Honscheid}, {Harding}, {Long}, {Lupton}, {Loomis}, {Anderson},
  {Annis}, {Bernardi}, {Bhardwaj}, {Bizyaev}, {Bolton}, {Brewington}, {Briggs},
  {Burles}, {Burns}, {Castander}, {Connolly}, {Davenport}, {Ebelke}, {Epps},
  {Feldman}, {Friedman}, {Frieman}, {Heckman}, {Hull}, {Knapp}, {Lawrence},
  {Loveday}, {Mannery}, {Malanushenko}, {Malanushenko}, {Merrelli}, {Muna},
  {Newman}, {Nichol}, {Oravetz}, {Pan}, {Pope}, {Ricketts}, {Shelden},
  {Sandford}, {Siegmund}, {Simmons}, {Smith}, {Snedden}, {Schneider},
  {SubbaRao}, {Tremonti}, {Waddell}, \& {York}}]{Smee13}
{Smee}, S.~A., {et~al.} 2013, \aj, 146, 32

\bibitem[{{York} {et~al.}(2000)}]{York00}
{York}, D.~G., {et~al.} 2000, \aj, 120, 1579

\end{thebibliography}

\end{document}